\documentclass[12pt, a4paper]{article}

\usepackage[
  margin=0.75in,
  headsep=10pt, 
]{geometry}

\usepackage{graphicx}
\usepackage{soul}
\usepackage{graphicx}
\usepackage{epstopdf, epsfig}
\usepackage{amsmath}
\usepackage[table]{xcolor}
\usepackage{subcaption}
\usepackage{tabularx}
\usepackage{soul}
\usepackage{makecell}
\usepackage{multirow}
\usepackage{breqn}
\usepackage{amssymb} 
\usepackage{booktabs}
\usepackage{dcolumn}
\usepackage{bm}
\usepackage[utf8]{inputenc}
\usepackage[T1]{fontenc}
\usepackage{mathptmx}
\usepackage{etoolbox}
\usepackage{hyperref}
\usepackage{subcaption}
\usepackage{ragged2e}
\usepackage[sort&compress]{natbib} 

\newcommand{\soptitle}{How does vortex dynamics help undulating bodies spread odor?}

\usepackage{xcolor} 

\begin{document}


\begin{center}
\Large \bf{\soptitle}
\vspace{0.1in}
\end{center}

\begin{center}
{Maham Kamran$^1$, Amirhossein Fardi$^1$, Chengyu Li$^2$, and Muhammad Saif Ullah Khalid$^{1\star}$}\\
\vspace{0.1in}
\end{center}
\begin{center}
$^1$Nature-Inspired Engineering Research Lab (NIERL), Department of Mechanical \& Mechatronics Engineering, Lakehead University, Thunder Bay, ON P7B5E1, Canada\\
$^2$Department of Mechanical and Aerospace Engineering, Case Western Reserve University, Cleveland, OH 44106, USA\\
\vspace{0.05in}
$^\star$\small{Corresponding Author, Email: mkhalid7@lakeheadu.ca}
\end{center}

\begin{abstract}

In this paper, we {examine} the coupling between odor dynamics and vortex dynamics around undulating bodies, with a focus on bio-inspired propulsion mechanisms. Utilizing computational fluid dynamics (CFD) simulations with an in-house Immersed-Boundary Method (IBM) solver, we investigate how different waveform patterns, specifically carangiform and anguilliform, influence the dispersion of chemical cues in both water and air environments. Our findings reveal that vortex dynamics significantly impact the overall trajectory of odor spots, although the alignment between odor spots and coherent flow structures is not always precise. We also evaluate the relative contributions of diffusion and convection in odor transport, showing that convection dominates in water, driven by higher Schmidt numbers, while diffusion plays a more prominent role alongside convection in air. Additionally, the anguilliform waveform generally produces stronger and farther-reaching chemical cues compared to carangiform swimmers. The critical roles of Strouhal number and Reynolds number in determining the efficiency of odor dispersion are also explained, offering insights that could enhance the design of more efficient, adaptive, and intelligent autonomous underwater vehicles (AUVs) by integrating sensory and hydrodynamic principles inspired by fish locomotion. 

\end{abstract}

\section{Introduction}
\label{sec:Intro}
Researchers always get fascinated by how nature {designs and adopts a variety of different effective strategies for dealing with} complex {systems}, particularly how animals like fish evolved highly efficient mechanisms for propulsion and navigation. The increasing use of {robotic technologies in aerial and aquatic environments for various purposes}, including environmental monitoring, search and rescue, and underwater exploration, {intensifies} research efforts to enhance their efficiency by drawing inspiration from natural designs. Fish are exceptional swimmers, and their ability to flex their bodies in specific ways enables them to achieve remarkable propulsion and maneuverability in aquatic environments. {Scientific and engineering community aims} to develop robots that can move through water {efficiently} by understanding the bio-mechanics and {hydrodynamics} of fish swimming \cite{triantafyllou2000hydrodynamics, sfakiotakis1999review}. Previously, research focused extensively on analyzing {propulsive mechanisms of fish} and {a} range of maneuvers achievable by manipulating factors{,} such as frequency and wavelength\cite{khalid2020flow}. Numerical simulations {are} pivotal in {examining and understanding} their dynamics, revealing that the wavelength of undulatory kinematics plays a crucial role in determining the hydrodynamic performance of fish \cite{zhang2022vortex}. {Recent research} studies {showed} that {undulating} with larger wavelengths \cite{khalid2021larger} significantly {enhanced} hydrodynamic thrust for {a carangiform swimmer}, allowing these fish to conserve energy at {an optimal wavelengths} while maintaining high swimming speeds \cite{anderson2001boundary}. {In another study, Khalid et al. \cite{khalid2021anguilliform} reported that anguilliform swimmers were able to attain better hydrodynamic performance metrics by performing the wavy motion at a smaller wavelength than their body-lengths. These findings demonstrate the diversity and effectiveness of different biological propulsive techniques in different marine animals}. Additionally, researchers investigated the influence of body shape and adaptive kinematics \cite{gupta2022anguilliform} to gain deeper insights into both propulsion and navigation capabilities. {Excellent reviews of these research investigations, laying} the groundwork for hydrodynamic designs of bio-inspired underwater robots {were presented by Fish \& Lauder \cite{fish2006passive, fish2017control}, Lauder \cite{lauder2015fish}, Fish \cite{fish2020bio}, Zhang et al. \cite{zhang2022physical}, and Raj and Thakur \cite{raj2016fish}}.

Fish {live and propel themselves} in {challenging marine environments} using a sophisticated combination of sensory modalities for underwater sensing, detection, and navigation \cite{montgomery2014sensory}. Researchers are keen to unravel the mysteries of aquatic locomotion and sensory perception to develop autonomous underwater vehicles (AUVs) that {can} mimic these natural abilities. Fish employ a diverse approach to gather information about their surroundings, utilizing visual inspection, olfaction (detection of chemical cues), acoustic perception, and sensitivity to pressure {fluctuations} through their lateral line system \cite{bleckmann2001lateral}. Visual inspection allows fish to perceive their environment, detect obstacles, and identify potential {preys} or predators \cite{domenici2011animal}. {Besides,} olfaction plays a crucial role in detecting chemical cues emitted by food sources, potential mates, or nearby predators, aiding in {migration} and foraging behaviors \cite{bunnell2011fecal}. Acoustic perception enables fish to sense disturbances in the water and communicate with conspecifics, contributing to social interactions and {strategies for avoiding} predators \cite{ladich2016peripheral}. Additionally, the sensitivity of fish scales {on their bodies} to pressure {changes} facilitates precise control of {directionality and} locomotion through water, enhancing their ability to perform complex maneuvers \cite{bleckmann2014central}.

Despite {many} advancements in understanding fish locomotion {during the last two decades}, the {important role and influence} of olfactory cues{, integrated with hydrodynamics, in} fish swimming remained {an entirely} unexplored aspect \cite{cox2008hydrodynamic, bronmark2012aquatic}. The interaction between hydrodynamics and chemosensory systems in fish is an emerging field of study that holds potential for significant discoveries \cite{hemmer2019genetic}. {To the authors' best knowledge, almost no studies were devoted} to {examine} the effects of chemical cues on fish locomotion \cite{webb2002control}. {Only one recent research study by Menzer et al. \cite{menzer2023multiphysics} addressed this important subject, who concluded that} olfactory chemoreception in a fish-like school and the associated hydrodynamic interactions played a crucial role in collective behavior and navigation. While it is well-established through research investigations from biologists \cite{cox2008hydrodynamic} that fish utilize chemical cues for {directional control during} navigation in water, this phenomenon was predominantly examined for {odor dynamics and aerodynamics of} flying insects \cite{lei2023wings, carde2021navigation}. However, a critical knowledge gap exists regarding the potential connection between fluid dynamics and odor dynamics around undulating bodies. In order to address this important {aspect related to biological propulsion in our present work}, we {employ} computational methods using {our} in-house solver, {developed based on Immersed-Boundary Method (IBM) and named as IBVortX \cite{farooq4874977accurate}}. The integration of these sensory and fluid dynamics insights {carries significant} potential to revolutionize the design of bio-inspired underwater robots \cite{bianchi2021bio}, {enabling them to become} more efficient, adaptive, and capable of performing complex tasks in challenging underwater environments. Mimicking the natural strategies of fish and leveraging their sensory capabilities, autonomous robotic systems can be made to respond intelligently to environmental cues \cite{marras2012fish} in diverse aquatic conditions, contributing to marine biology, environmental conservation, and oceanographic exploration. {Therefore, combining vortex dynamics with odor dynamics around undulating bodies fundamentally sets up the novelty and the main objective of our current work.}

{In order to further elaborate the research questions connected with the primary aim of our work, we address the following elements here: (i) how does odor dynamics couple with vortex dynamics around undulating bodies in water? (ii) modeling the undualting body as a source of odor, which waveform, carangiform or anguilliform, sends strong chemical cues in the wake? (iii) how does diffusion and convection of odor comparatively behave in a fluid medium for bio-inspired propulsion? (iv) how does flow conditions and kinematic parameters influence the apparently interconnected dynamics of fluid flow and transport of chemical cues? and (v) although undulation of thin bodies relates more to fish swimming as its model, how does the nature of a fluid medium, e.g., water and air, affects the diffusion and convection processes of odor? It is important to highlight that our present work is the first study that involves the computational modeling of both diffusion and convection phenomena in determining the unsteady behavior and transport of a chemical specie at both high and low values of Schmidt numbers \cite{li2018balance, lei2023wings}. 

There are different classifications of marine species based on their physiology and kinematics \cite{sfakiotakis1999review}. These include anguilliform, carangiform, sub-caragiform, thunniform, and ostraciiform. Our current work is only focussed on two primary modes, including carangiform and anguilliform. We consider viscous and transitional flow regimes, defined through Reynolds numbers ($\mbox{Re}$) of 500, 1000, and 5000 \cite{khalid2020flow} along with Strouhal number, denoted as $f^\ast$, ranging from 0.2 to 0.6. Here, we characterize fluid media, water and air, through} {Schmidt number, which is the ratio between kinematic viscosity ($\nu$) of a fluid and diffusivity ($D$) of an odor}, as provided in Table~\ref{performance_parameters}. Despite the fact that fish primarily inhabit aquatic environments, testing the solver's robustness in both water and air is essential for comprehensive validation and verification. 

\section{Computational Methodology}
\label{sec:Num_Method}

\begin{figure}[htbp]
{\includegraphics[width=1.0\textwidth]{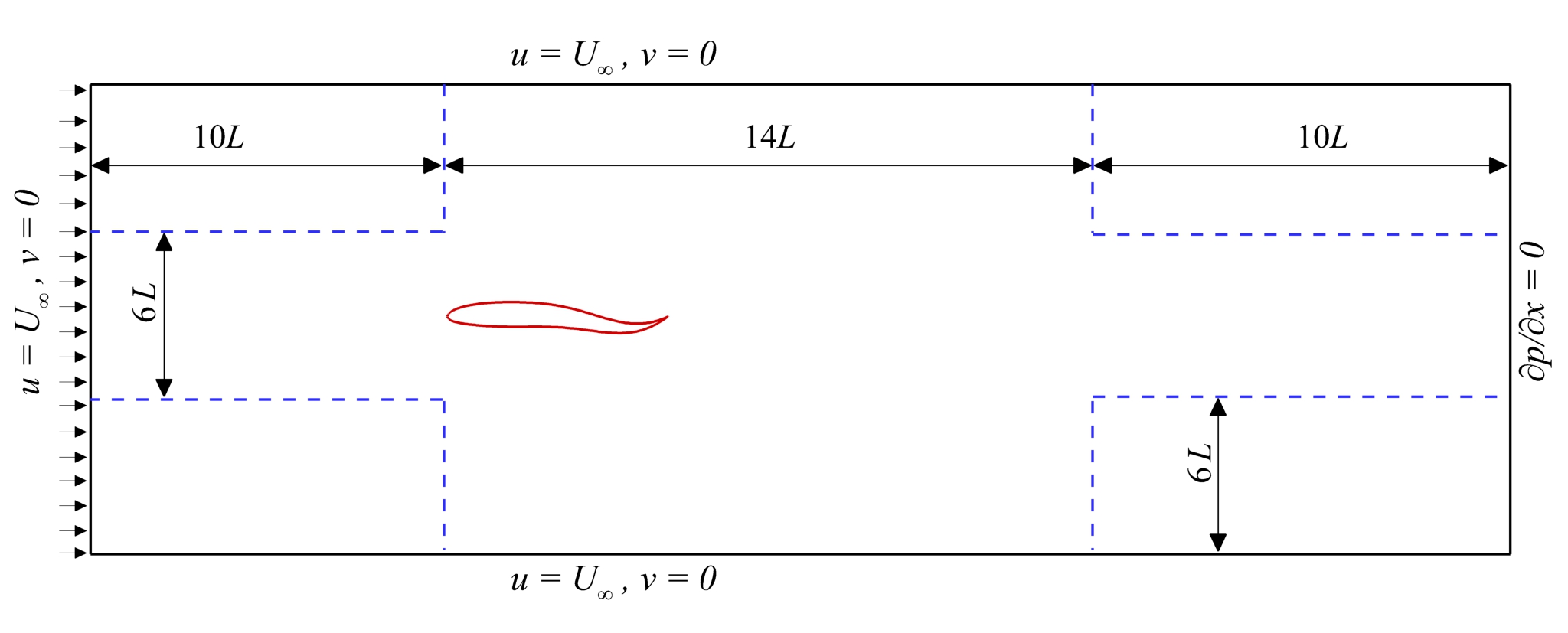}}
\caption{Flow domain and the boundary conditions}
\label{fig:flow_domain}
\end{figure}

{In this study, we use the NACA0012 foil to model the bodies of swimmers, with the chord representing a swimmer's spine during static equilibrium. We consider two types of wavy kinematic modes: anguilliform and carangiform. Anguilliform swimmers, such as eels, move by undulating a large portion of their bodies, while carangiform swimmers, like certain fish species, have prominent caudal fins attached to their bodies \cite{lauder2005hydrodynamics}. The amplitude of the carangiform profile, where the chord represents a fish’s backbone, is described by the following relation \cite{khalid2016hydrodynamics, khalid2020flow, khalid2021larger}:

\begin{align}
A(\frac{x}{L}) &= 0.02 - 0.0825(\frac{x}{L}) + 0.1625(\frac{x}{L})^2; 0 < \frac{x}{L} < 1
\label{eq:carangiform}
\end{align}

\noindent where $x$ denotes the stream-wise coordinate of each node used to discretize the model swimmer, and $A(x/L)$ shows the local amplitude at a given spatial position along the swimmer's body, nondimensionalized by its total length ($L$). This body length is equal to the chord of the foil for our $\mbox{2D}$ simulations. Here, The coefficients are calculated based on the data provided for a steadily swimming saithe fish, which is a carangiform swimmer \cite{videler1993fish} with local amplitudes of $A(0)=0.02$, $A(0.2)=0.01$, and $A(1.0)=0.10$. Limiting the maximum amplitude of the trailing-edge to $0.10$ for the two swimmers, we define the amplitude envelop of anguilliform kinematics using the following relation \citep{maertens2017optimal,khalid2020flow,khalid2021anguilliform}:

\begin{align}
    A(\frac{x}{L}) &= 0.0367 + 0.0323(\frac{x}{L}) + 0.0310{(\frac{x}{L})}^2; \quad 0 < x/L < 1
\label{eq:amp}
\end{align}

Using $f$ being the oscillation frequency and $t$ as the time, the undulatory motion in both the cases is modeled by the following mathematical form: 

\begin{align}
    y(\frac{x}{L}) &= A(\frac{x}{L}) \sin[2\pi({\frac{x}{\lambda}}-ft)]
\label{eq:motion}
\end{align}

The swimmer is positioned in a rectangular virtual tunnel, as illustrated in Fig.~\ref{fig:flow_domain}, with dimensions of the flow domain set to as ${34L}\times{18L}$. Please note that the figure is not intended to be according to a scale but rather to highlight the geometric details.

}

{We conduct two-dimensional (2D) numerical simulations of the swimmers, disregarding the effects of three-dimensional flow. This approach serves as an efficient computational strategy for capturing key aspects of swimming dynamics and flow physics \cite{khalid2016hydrodynamics, gazzola2014scaling}. The mathematical model for fluid flow is based on the following non-dimensional forms of the continuity and incompressible Navier-Stokes equations.

\begin{align}
\frac{\partial u_j}{\partial x_j} &=0\\
\frac{\partial u_i}{\partial t} + {u_j}\frac{\partial}{\partial
x_j}({u_i}) &= -\frac{1}{\rho}\frac{\partial p}{\partial
x_i}+ \frac{1}{Re}\frac{{\partial^2}u_i}{{\partial x_j}{\partial x_j}} + f_b
\label{eqn:CartCont}
\end{align}


\noindent where $\{i,j\}=\{1,2,3\}$, $x_i$ shows Cartesian directions, $u_i$ denotes the Cartesian components of the fluid velocity, $p$ is pressure, and $\mbox{Re}$ represents Reynolds number. For the current numerical formulation, the term $f_b$ shows a discrete forcing term.    

We solve the governing mathematical model for fluid flows using a sharp-interface immersed boundary method on a non-uniform Cartesian grid, with radial-basis functions as the interpolation scheme to precisely identify the immersed bodies \cite{farooq4874977accurate}. We use a central difference scheme for spatial discretization to approximate the diffusion term, while the convection term is discretized using the Quadratic Upstream Interpolation for Convective Kinematics (QUICK) scheme. The integration with respect to time is carried out through a fractional-step method, ensuring second-order accuracy in both time and space. The prescribed wavy kinematics are applied as a boundary condition on the swimmer's body, enforced on immersed bodies through a ghost-cell technique \cite{mittal2008versatile,farooq4874977accurate} suitable for both rigid and flexible structures. Further details on this fully parallelized solver and its application to various bio-inspired fluid flow problems can be found in the work of Farooq et al \cite{farooq4874977accurate}. Neumann boundary conditions are applied at the far-field boundaries, except for the left-sided inlet boundary of the domain, where Dirichlet conditions are used for the in-flow. The dashed lines in Fig.~\ref{fig:flow_domain} mark the boundaries of the zone with high mesh density to accurately capture the flow characteristics around the body and its wake.
}
Odor concentration refers to the chemical cues released in a fluid medium, which can indicate the presence of prey, predators, or potential mates, thus aiding in chemical sensing for navigation, foraging, and communication. These chemical signals consist of various molecules, such as amino acids, pheromones, and metabolic waste products, which diffuse through aerial and aquatic environments. {After computing the velocity field $u_i$, the odor transport equation is solved to determine the instantaneous odor concentration field in the whole computational domain. The governing unsteady equation for the convection and diffusion of the odorant is provided below:

\begin{align}
\frac{\partial C}{\partial t} + {u_i}\frac{\partial{C}}{\partial
x_i} &= {D}\frac{{\partial^2}C}{{\partial x_i}{\partial x_i}} 
\label{eqn:odor_transport}
\end{align}

\noindent Where $C$ is the odor concentration, nondimensionalized by the source odor concentration at the body surface of swimmers, and $D$ is the diffusivity of odor. To numerically approximate the temporal (first term on the left side), convective (second term of the left side), and diffusion terms (the only term on the right side) in Eq.~\ref{eqn:odor_transport} using the same computational schemes for temporal and diffusion terms that we employ to compute the analogous terms in the Navier-Stokes Equation (Eq.~\ref{eqn:CartCont}). We solve the convective term using the upwind scheme. We explain the role of convection and diffusion processes in transport of odor in the flow field using the respective terms in Eq.~\ref{eqn:odor_transport}. Our robust approach enables us to perform multi-physics computational simulations, involving fluid-structure-chemical interactions, with very high values of Schmidt number accurately and efficiently. 
}

\subsection{{Validation \& Verification}}
\begin{figure}[htbp]
    \centering
    \subfloat[\centering]{{\includegraphics[width=8cm]{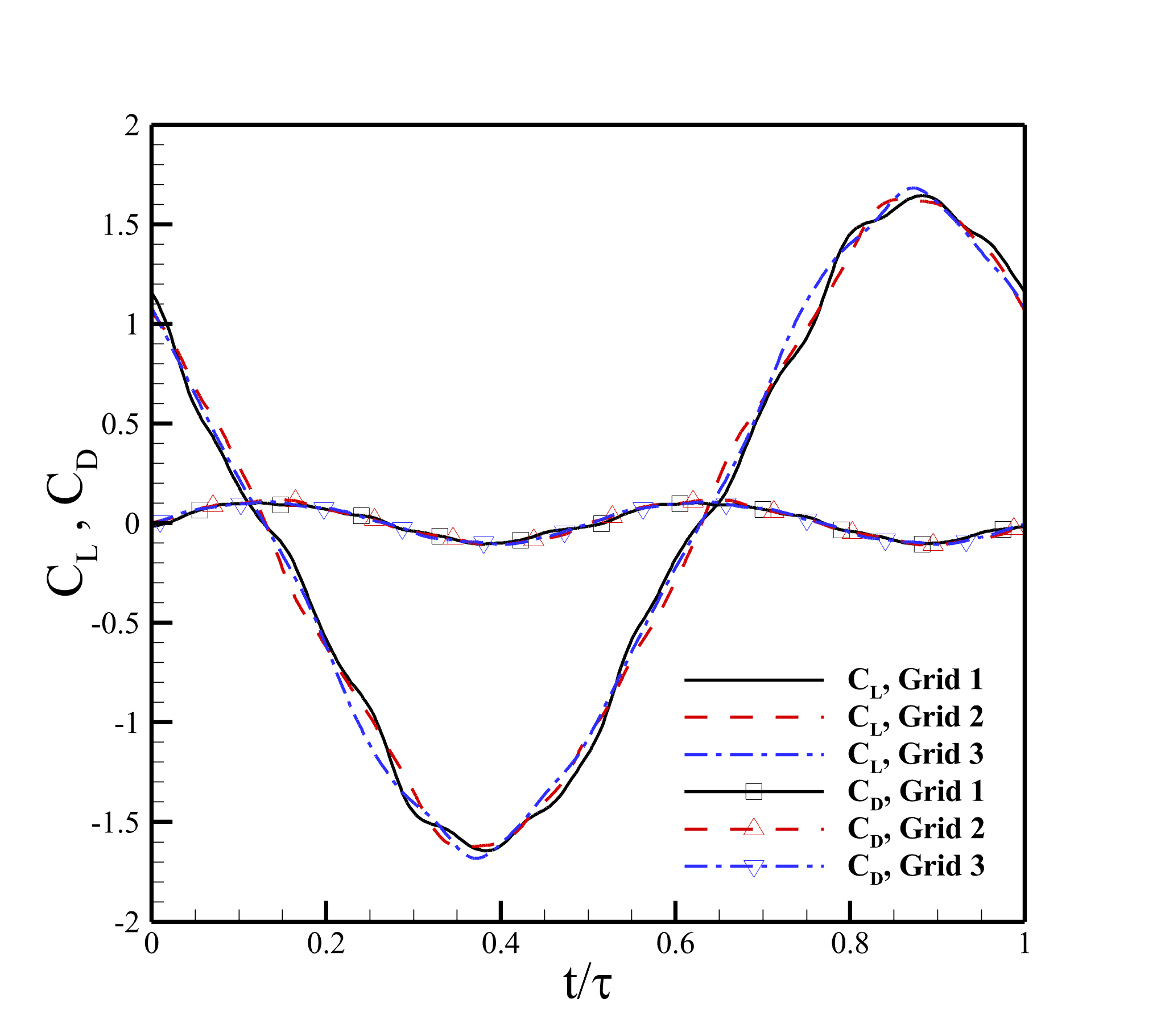} }}%
    \qquad
    \subfloat[\centering]{{\includegraphics[width=8cm]{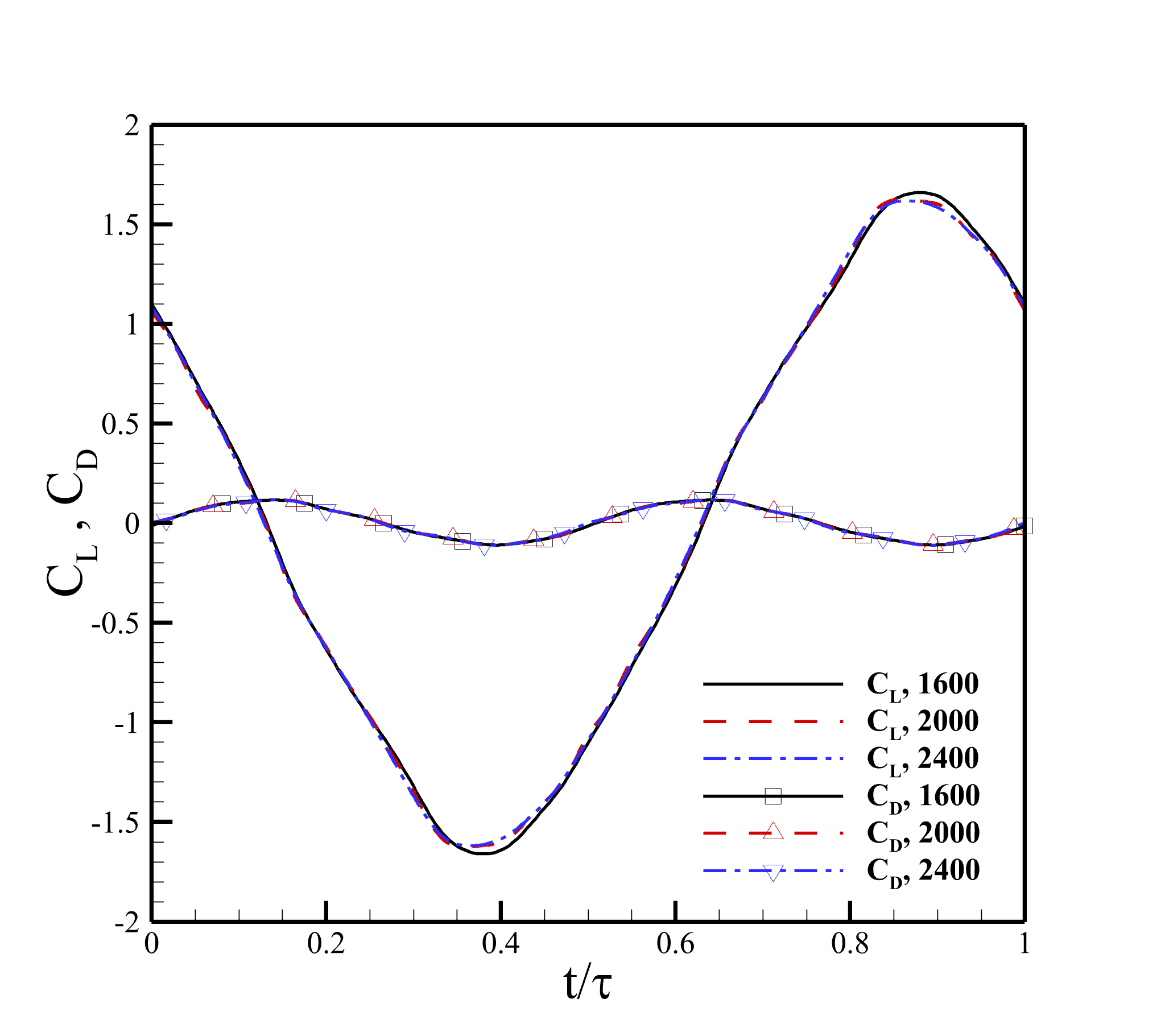} }}%
       \caption{\small\justifying Results for convergence of (a) grid size and (b) time-step size}%
    \label{Results for convergence}%
\end{figure}

{Before performing our multi-physics simulations for the current work, we carry out an extensive sets of grid-independence and time-step convergences studies. For this purpose, we employ a {NACA-0012} foil by prescribing a carangiform waveform over its surface \cite{khalid2020flow}, characterized by an undulation wavelength ($\lambda$) of $1.05$. Please note that $\lambda$ is nondimensionalized by the chord-length of the foil, and the Reynolds number defined by $\mbox{Re}={U_\infty}{L}/\nu$ is set as $10^3$. Here, $U_\infty$ and $L$ denote the free-stream velocity of the flow and chord-length of the foil, respectively. We conduct these simulations for a Strouhal number $\mbox{St}=0.40$. First, we complete the grid-independence test using three different grid configurations using a time-step ($\Delta{t}$) according to $2000$ time-steps per oscillation cycle. Grid 1, grid 2, and grid 3 have sizes of $1597\times1135$, $1747\times1201$, and $1975\times1273$, respectively. Figure~\ref{Results for convergence}a exhibits comparisons between the instantaneous lift and drag coefficients, represented by $C_L$ and $C_D$ of the undulating foil for its one undulation cycle, where $\tau$ is the time-period of the undulation. Here, the profiles of the force coefficients qualitatively look very similar for the three grids during the $15^{\text{th}}$ cycle, where the solutions become steady-state within $5-6$ oscillation cycles. } 

{In order to make an adequate choice of the grid for our next simulations and to better evaluate the convergence of our results, we quantify the difference between the three solutions through a parameter called the standard error of estimate ($SE_e$). It is mathematically defined as:}
\[
SE_e = \sqrt{\frac{\sum (Y_{\text{obs}} - Y_{\text{pred}})^2}{n - p}}
\]
where \( Y_{\text{obs}} \) represents the observed values, \( Y_{\text{pred}} \) shows the predicted values, \( n \) is the number of observations, and \( p \) is the number of parameters. {Choosing the results obtained from Grid 3 as the reference ones {{($Y_{\text{pred}}$)}}, Table~\ref{Comparison of Grid and Time-step independence study results} clearly reflects smaller differences between the results obtained from Grid 2 and Grid 3 for both $C_L$ and $C_D$. Therefore, we proceed with Grid 2 for our next set of simulations. For the time-step independence study, we choose three values of $\Delta{t}$, $\Delta{t}_1$, $\Delta{t}_2$, and $\Delta{t}_3$, according to $1600$, $2000$, and $2400$ time-steps, respectively, in each oscillation cycle. The results in terms of $C_L$ and $C_D$ are presented in Fig.~\ref{Results for convergence}b, where we find the three profiles of each force coefficient to be very similar. In order to further ensure the convergence of $\Delta{t}$ using the smallest $\Delta{t}$ as the reference values (Time-step 3 in this case), values of $SE_e$ are shown in Table~\ref{Comparison of Grid and Time-step independence study results}. It is clear that the difference in results between the simulations for $\Delta{t}_2$ and $\Delta{t}_3$ is reduced compared to those from $\Delta{t}_1$ and $\Delta{t}_3$. Hence, the remaining simulations for this work are performed using $2,000$ time-steps per undulation cycle. For validation of our computational methodology, we refer the readers to the very recent work of Farooq et al. \cite{farooq4874977accurate}.}

\begin{table}[htbp]
\centering 
\caption{\small Comparison of {values of $SE_e$ for grid-convergence and time-step independence tests}}
\vspace{-25pt}
\resizebox{\textwidth}{!}{%
\begin{tabular}{@{}cccccccc@{}}
\multicolumn{1}{l}{} & \multicolumn{1}{l}{} & \multicolumn{1}{l}{} & \multicolumn{1}{l}{} & \multicolumn{1}{l}{} \\ \toprule\toprule
& \multicolumn{2}{c}{Grid-independence Tests}  &&  \multicolumn{2}{c}{Time-step Convergence Tests} \\ \cline{2-3} \cline{5-6}
\textbf{} & Grid 1 \& Grid 3 & Grid 2 \& Grid 3 &&  $\Delta{t}_1$ \& $\Delta{t}_3$ &  $\Delta{t}_2$ \& $\Delta{t}_3$ \\ \hline
Drag coefficient ($C_D$) & 0.8221  &  0.5838 &&    0.4393 &  0.2475 \\
Lift coefficient ($C_L$) & 6.0575 &  5.8059 & &   2.5258 &  1.1156 \\ \bottomrule \bottomrule
\end{tabular}%
}
\label{Comparison of Grid and Time-step independence study results}
\end{table}

\begin{figure}[htbp]
\centering
{\includegraphics[width=0.8\textwidth]{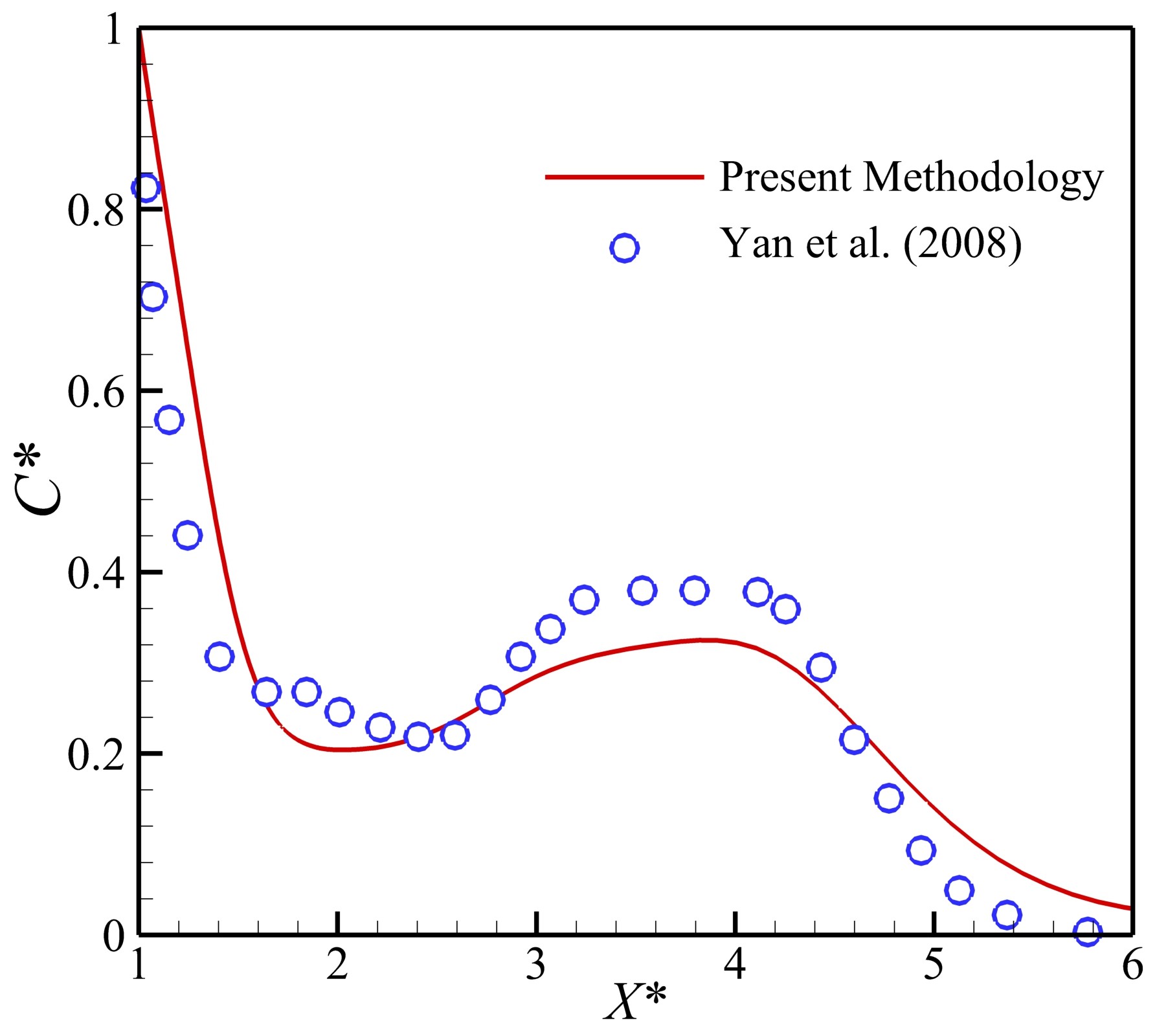}}
\caption{Comparison of our results with those of Yan et al. \cite{yan2008numerical} from the transport equation}
\label{fig:valid_odor}
\end{figure}

{To validate the accurate functionality of the transport equation (Eq.~\ref{eqn:odor_transport}), we adopt the route suggested by Lei et al. \cite{lei2021navigation}. The structure of the advection-diffusion equation for transport of an odor is the same as that of the energy equation. In Fig.\ref{fig:valid_odor}, we present the comparison of our results (in terms of $C^\ast$) for a $\mbox{2D}$ flow around a rotating cylinder with the results obtained by Yan et al. \cite{yan2008numerical}, who used their computational tool developed based on Lattice-Boltzzman method. These results are obtained for $\mbox{Re}=200$ and a reduced frequency of $0.5$. A good match between the two instaneous profiles of the dependent parameters versus $X^\ast$ (the x-coordinate nondimensionalized by radius of the cylinder) demonstrates the effectiveness of our solver. 
}

\section{Results and Discussion}
{To find answers of specific research questions, defined in section~\ref{sec:Intro}, related to coupled vortex dynamics and transport of chemical cues, we perform simulations for flows and odor dynamics around $\mbox{2D}$ undulating bodies. The details of the governing flow, kinematic, and odor-related parameters are provided in Table~\ref{performance_parameters}. We run these simulations using three different values of $\mbox{Re}$, indicating viscous and transitional flow regimes \cite{khalid2020flow}. Two different wave-forms, anguilliform and carangiform, are prescribed to define undulation of the body \cite{khalid2020flow} at $f^\ast$, ranging from 0.2 to 0.6 with a gap of 0.05, as marine animals mostly swim in this range \cite{khalid2020flow}. We define two different fluid media using $Sc$ as $340$ (for water) and $0.7$ (for air). These specifications make the total number of simulations as  108 for our present work. It is important to mention that we employ $\lambda=0.75$ for the anguilliform kinematic mode \cite{khalid2021anguilliform} and $\lambda=1.05$ for the carangiform one \cite{khalid2021larger}.}

\begin{table}[htbp]
\centering
\caption{\small {Specifications of governing parameters}}
\setlength{\tabcolsep}{50pt} 
\begin{tabular}{ll}
\hline
\hline
Parameters           & Specifications              \\ \hline
Geometry              & NACA-0012                    \\
Undulatory kinematics & Anguilliform and Carangiform \\
Re                    & 500, 1000, and 5000           \\
Sc                    & Water (340) and Air (0.7)     \\
$f^\ast$                 & 0.2 - 0.6                    \\
$\lambda$             & 0.75 and 1.05                \\ \hline
\hline
\end{tabular}
\label{performance_parameters}
\end{table}

{We begin the discussion on our results using} qualitative and quantitative analyses of hydrodynamic performance {metrics} for both anguilliform and carangiform swimmers.

\begin{figure}[h!]
    \centering
    \begin{minipage}[b]{0.45\linewidth}
        \centering
        \includegraphics[width=\linewidth]{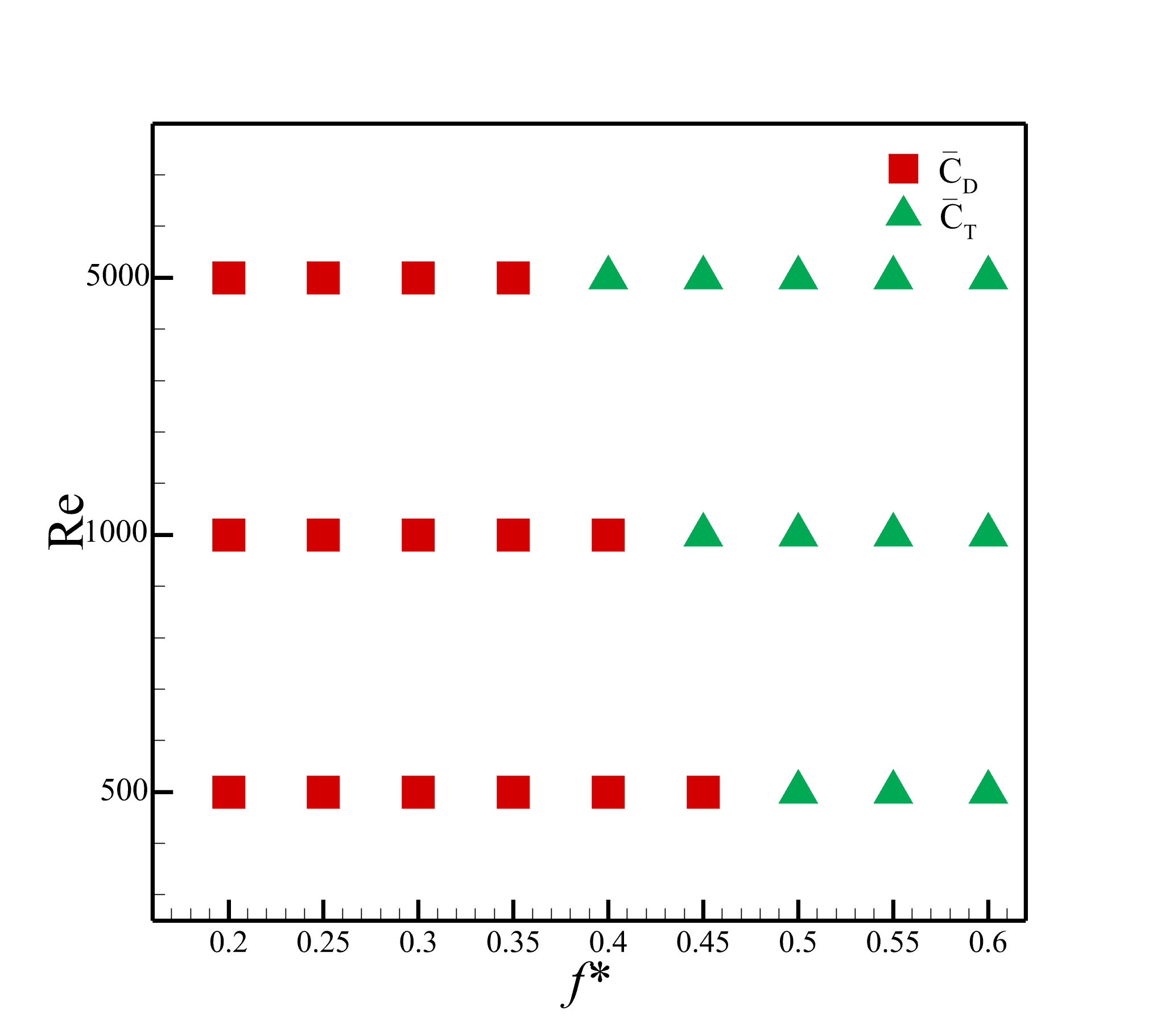}
        \caption*{(a)}
    \end{minipage}
    \hfill
    \begin{minipage}[b]{0.45\linewidth}
        \centering
        \includegraphics[width=\linewidth]{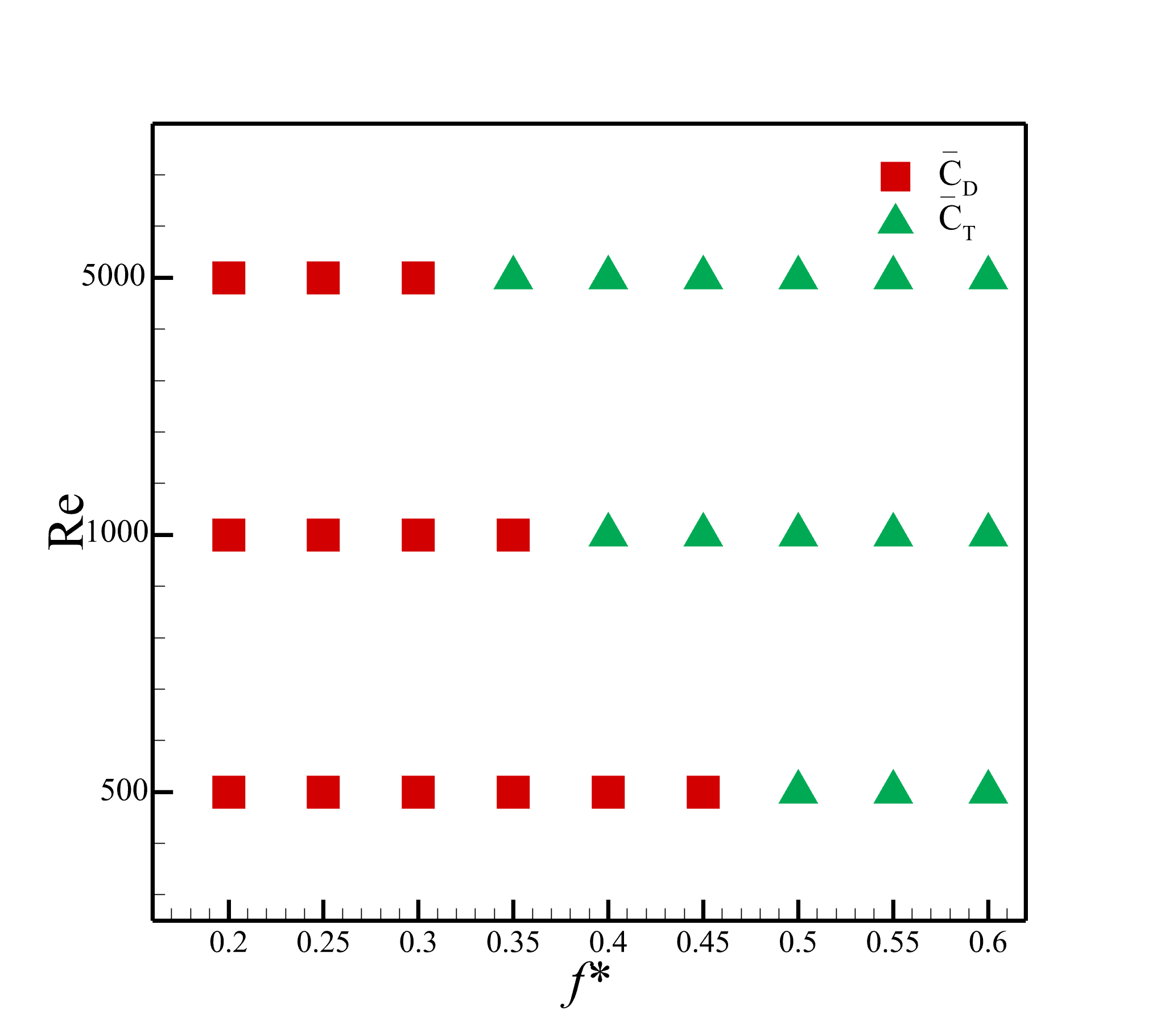}
        \caption*{(b)}
    \end{minipage}
    \caption{\small\justifying {Phase maps illustrating the flow regimes and their transitions for both undulating swimmers, (a) and (b) correspond to anguilliform and carangiform swimmers, respectively at Re = 500, 1000 and 5000.}}
    \label{General Lift and Drag trend}
\end{figure}

\begin{figure}[h!]
    \centering
    \begin{minipage}[b]{0.45\linewidth}
        \centering
        \includegraphics[width=\linewidth]{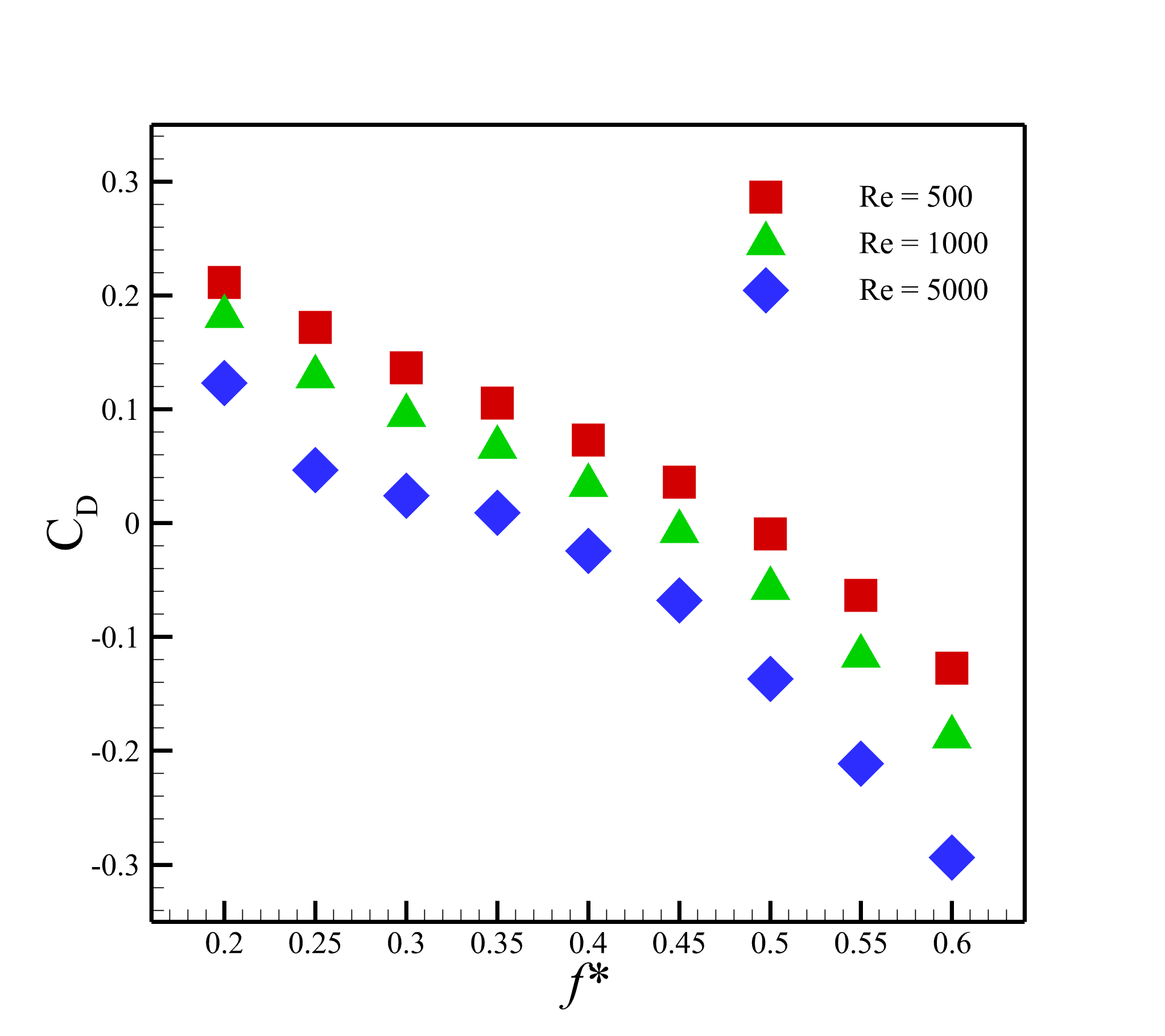}
        \caption*{(a)}
    \end{minipage}
    \hfill
    \begin{minipage}[b]{0.45\linewidth}
        \centering
        \includegraphics[width=\linewidth]{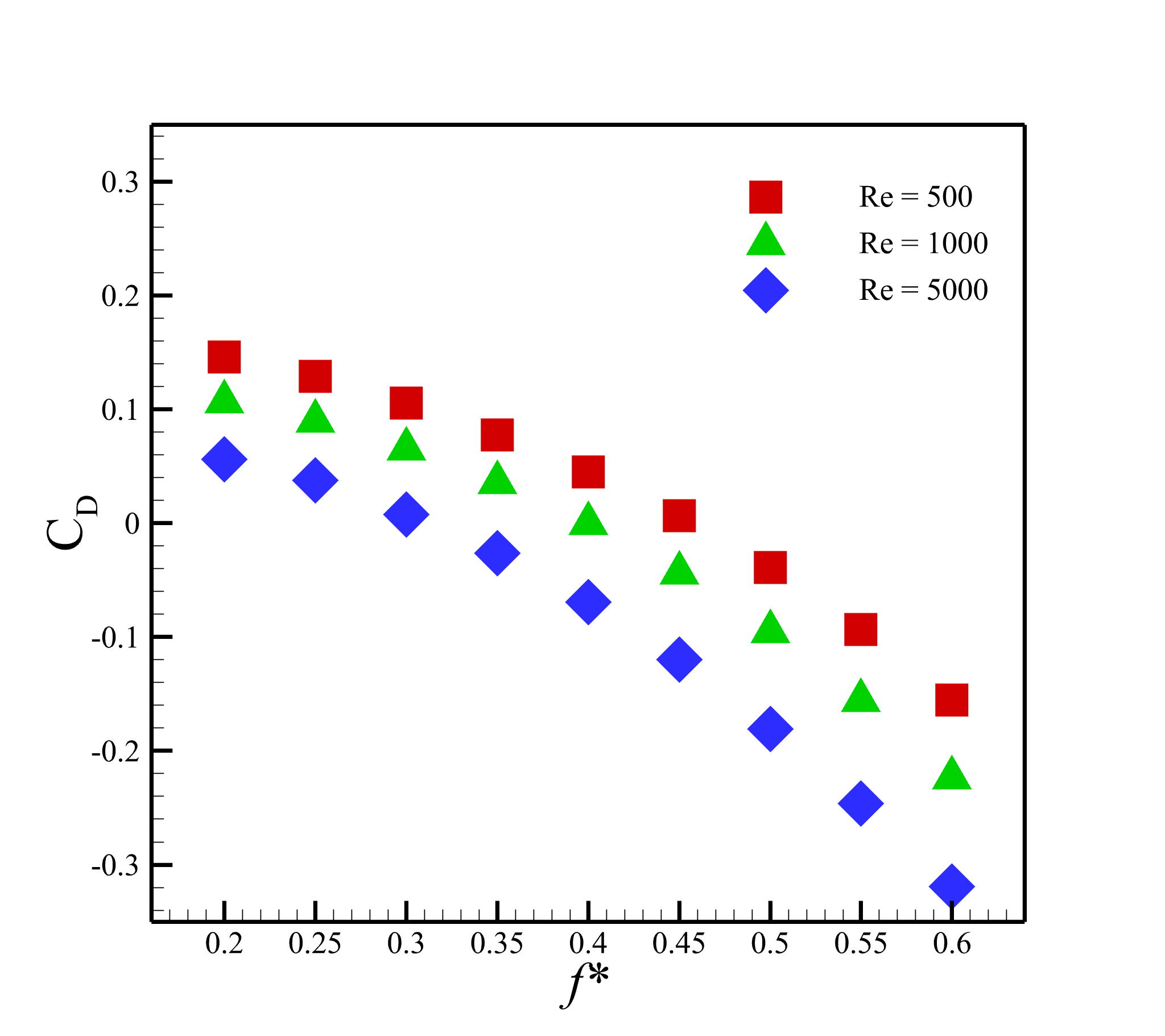}
        \caption*{(b)}
    \end{minipage}
    \caption{\small\justifying Average drag coefficient values with different $f^\ast$ at Re = 500, 1000 and 5000 where (a) and (b) correspond to anguilliform and carangiform swimmers, respectively.}
    \label{Lift and Drag values}
\end{figure}

{In the form of phase maps, Fig.~\ref{General Lift and Drag trend} provides distinct combinations of $\mbox{Re}$ and $f^\ast$ that produce overall drag and thrust forces. The information corroborates with the findings of Khalid et al. \cite{khalid2020flow} in terms of showing the production of thrust at $f^\ast \ge 0.5$. This critical value of Strouhal number decreases, as we increase $\mbox{Re}$, slowly for anguilliform swimmer and more aggressively for the carangiform one. Additionally, Fig.~\ref{Lift and Drag values} presents data for time-averaged $C_D$ versus $f^\ast$ for both swimming modes at the three values of $\mbox{Re}$. The negative values of $C_D$ corresponds to the generation of thrust here. In a way, certain combinations of $\mbox{Re}$ and $f^\ast$, demonstrating ${C_D}{\sim}0$, exhibit flow and kinematic parameters for self-propulsion or free swimming conditions for the two kinds of swimmers.}

\begin{figure}[htbp]
    \centering
    \begin{minipage}[b]{1\linewidth}
        \centering
        \includegraphics[width=\linewidth]{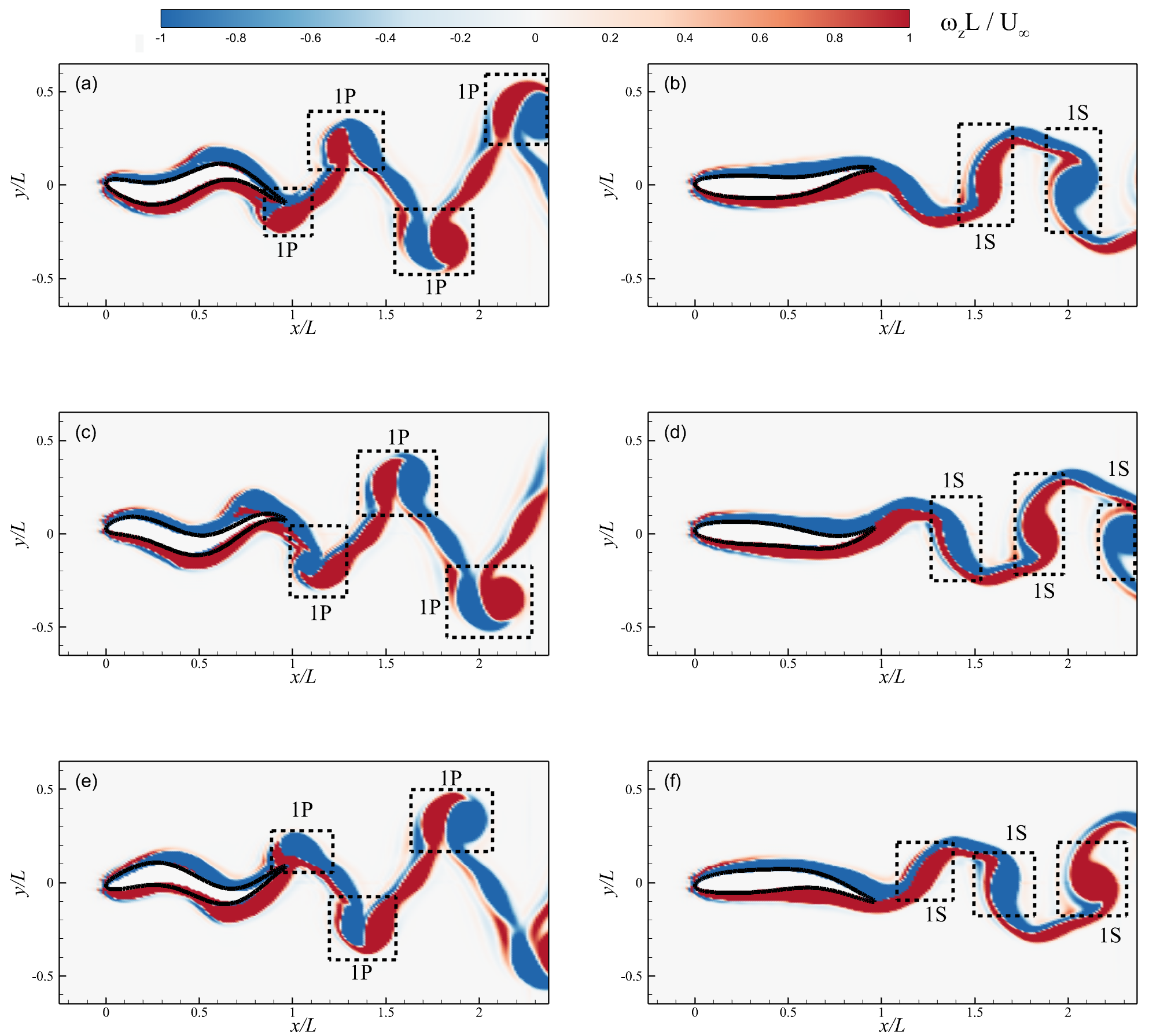}
    \end{minipage}
    \caption{\small\justifying {Vortex configuration patterns at different time instants during the undulation cycle. The first row corresponds to $t/\tau = 0$, the second row to $t/\tau = 0.50$, and the third row to $t/\tau = 1.00$. Panels (a), (c), and (e) show vortex patterns for anguilliform swimmers; panels (b), (d), and (f) for carangiform swimmers.}}
    \label{2s2p_Re_5000}
\end{figure}

{To understand the dependence of transport of odor on vortex dynamics around and in the wake of undulating bodies, there appears two direct routes: (i) we consider the vortex wake configurations through the classifications of von-Karman vortex street, neutral wake, and reverse von-Karman vortex street \cite{godoy2008transitions}, and (ii) we carefully observe the classifications of the wake in terms of $S$, and $P$ wakes \cite{ali2021flow}, where $S$ and $P$ are the terms for a single vortex and a pair of vortices, respectively.

\begin{figure}[htbp]
    \centering
    \begin{minipage}[b]{0.45\linewidth}
        \centering
        \includegraphics[width=\linewidth]{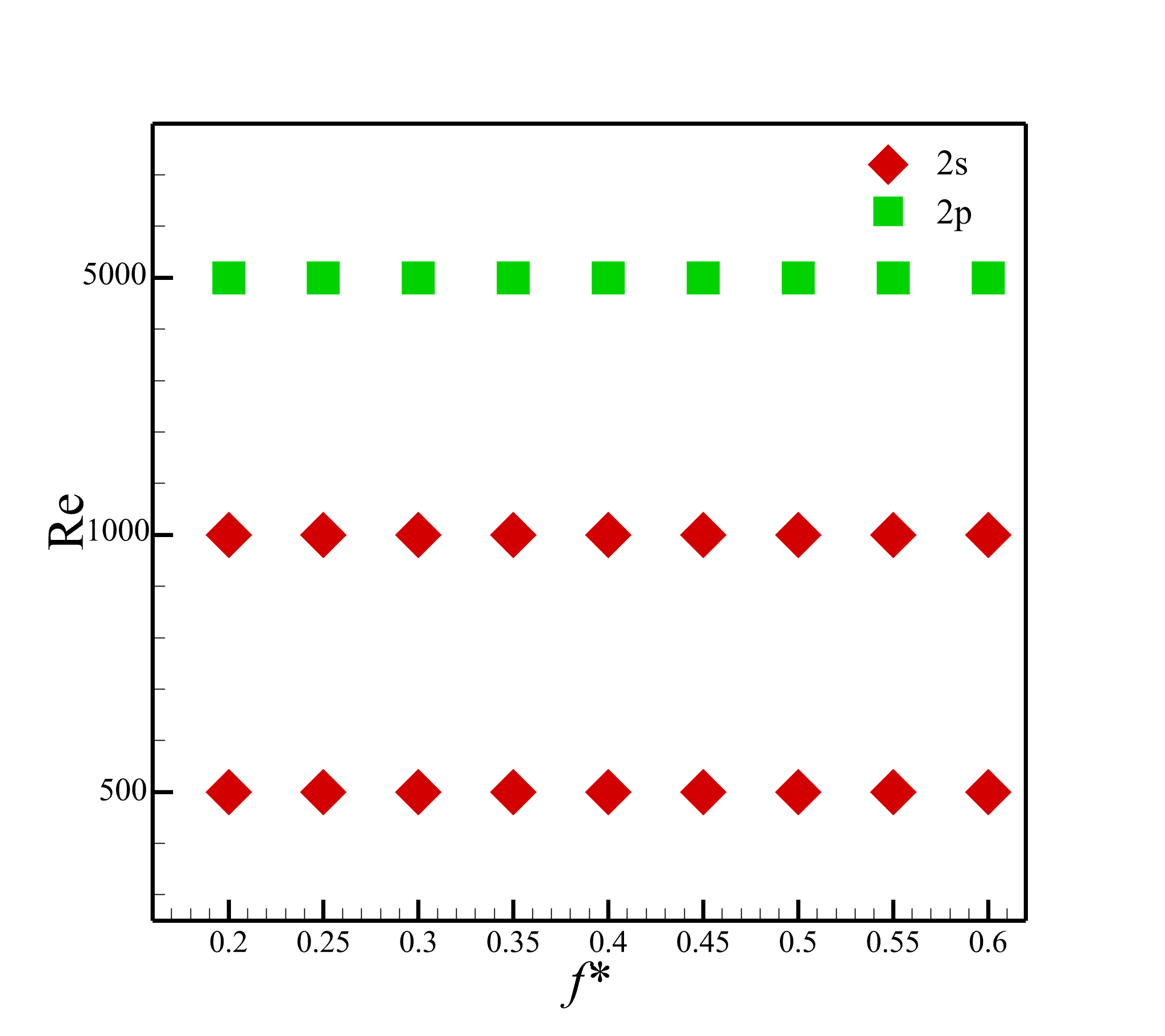}
        \caption*{(a)}
    \end{minipage}
    \hfill
    \begin{minipage}[b]{0.45\linewidth}
        \centering
        \includegraphics[width=\linewidth]{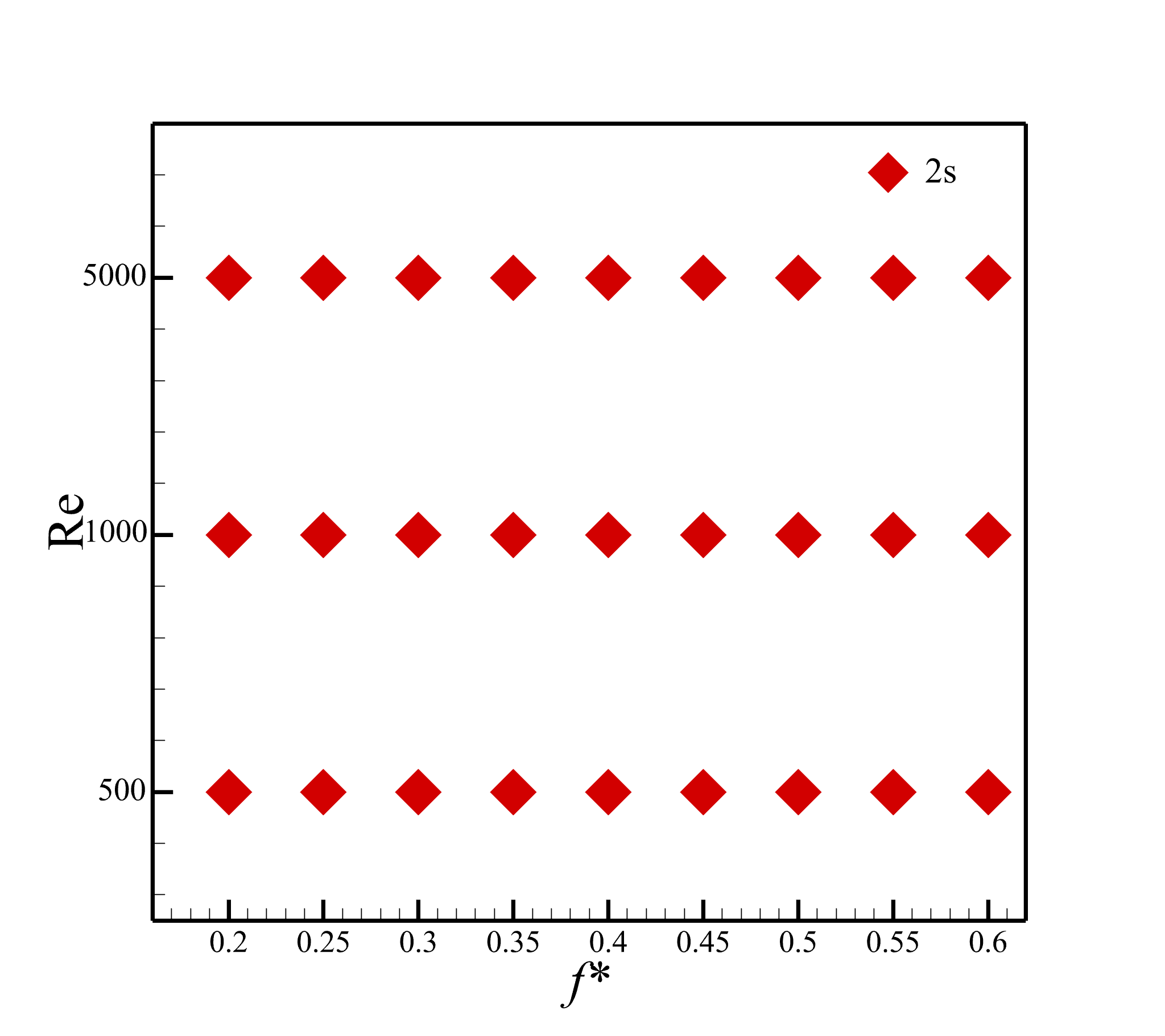}
        \caption*{(b)}
    \end{minipage}
    \caption{\small\justifying Vortex configuration at Re = 500, 1000 and 5000 where (a) and (b) correspond to anguilliform and carangiform swimmers, respectively.}
    \label{Vortex patterns}
\end{figure}

The 2S pattern involves the shedding of a single vortex in each half oscillation cycle, whereas the 2P pattern involves shedding of a pair of counter-rotating vortices in each half undulation cycle. The later choice seems more viable, because our purpose here is not to connect the odor dynamics with the production of drag or thrust forces, but to explain the transport of chemical cues, apparently dependent on formation, shedding, and dynamics of vortices. Therefore, we adopt the second route to further explain our analysis. As evident from Figs.~\ref{Vortex patterns}a and \ref{Vortex patterns}b, most of the parametric space corresponding to both anguilliform and carangiform swimmers consistently indicates the production of 2S wake, whereas only the anguilliform swimmer exhibits its wake transitioned to 2P patterns at $\mbox{Re}=5,000$. Using the out-of-plane vorticity ($\omega_z$), we present these two distinct patterns of wakes in Figs.~\ref{2s2p_Re_5000}a \ref{2s2p_Re_5000}b for the anguilliform and carangiform swimmers, respectively for $\mbox{Re}=5,000$ and $f^\ast=0.2$, observed at three time instants: $t/\tau = 0$, $t/\tau = 0.50$, and $t/\tau = 1.00$. The Carangiform swimmer shed only a single primary vortex in each half oscillation cycle, whereas the anguilliform one sheds a pair of counter-rotating vortices in each half oscillation cycle, {as explained by the annotations in these contour plots}.} 

{Because dynamics of the chemical cues are expected to be significantly influenced by the vortices generated by the undulating bodies,} the next step is to investigate whether the distribution of odor follows the same patterns as the vortex dynamics. To explore their relation, we {analyze flow and odor fields under various conditions according to the governing parameters in this study. Figures~\ref{ov_angulli} and \ref{ov_carangi} provides side-by-side comparative contour plots for the transport of vorticity and odor produced by anguilliform and carangiform swimmers, respectively. The plots in Fig.~\ref{ov_angulli}a$-$~\ref{ov_angulli}d and Figs.~\ref{ov_carangi}a$-$\ref{ov_carangi}d represent the data for $\mbox{Re}=500$ at $f^\ast=0.2$ and $f^\ast=0.6$. Similarly, the data presented in Figs.~\ref{ov_angulli}a$-$\ref{ov_angulli}d and Figs.~\ref{ov_carangi}a$-$\ref{ov_carangi}d correspond to $\mbox{Re}=5,000$ at the same Strouhal frequencies. The first observation from the comparison of all the plots relates to the odor fields strongly linked with vortex dynamics in the the wake. It is apparent that odor spots primarily seem to follow the trajectory of the vortices. 

\begin{figure}[htbp]
    \centering
    \begin{minipage}[b]{1\linewidth}
        \centering
        \includegraphics[width=\linewidth]{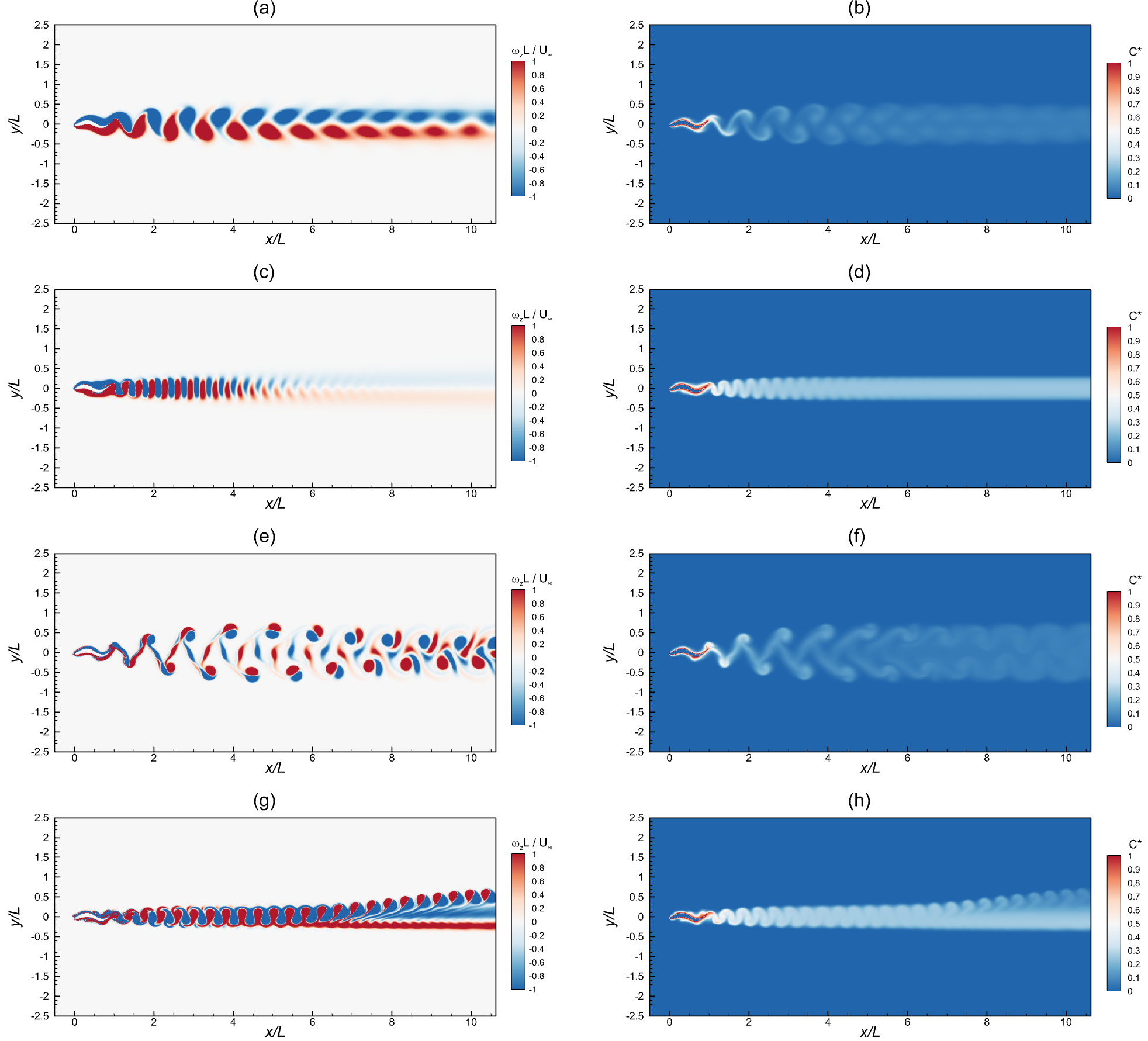}
    \end{minipage}
    \caption{\small\justifying Vortex and odor concentration for anguilliform swimmers, with the first and second columns corresponding to vorticity contours and odor concentration, respectively. Panels (a), (b) and (c), (d) display scenarios with a Re = 500 and $f^\ast$ = 0.2 and 0.6, respectively. Panels (e), (f) and (g), (h) display scenarios with a Re = 5000 and $f^\ast$ = 0.2 and 0.6, respectively.}
    \label{ov_angulli}
\end{figure}

\begin{figure}[h!]
    \centering
    \begin{minipage}[b]{1\linewidth}
        \centering
        \includegraphics[width=\linewidth]{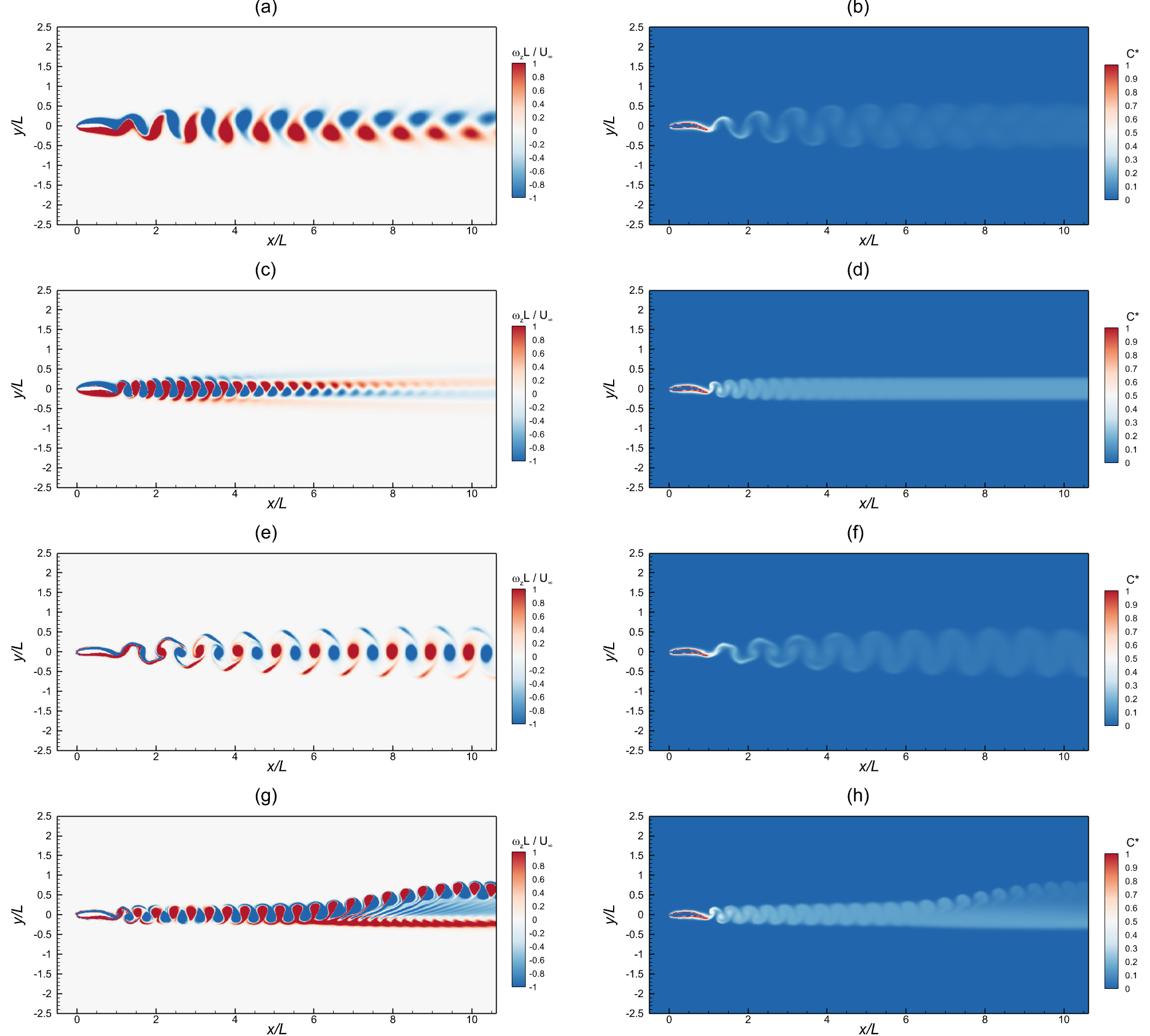}
    \end{minipage}
    \caption{\small\justifying Vortex and odor concentration for carangiform swimmers, with the first and second columns corresponding to vorticity contours and odor concentration, respectively. Panels (a), (b) and (c), (d) display scenarios with a Re = 500 and $f^\ast$ = 0.2 and 0.6, respectively. Panels (e), (f) and (g), (h) display scenarios with a Re = 5000 and $f^\ast$ = 0.2 and 0.6, respectively.}
    \label{ov_carangi}
\end{figure}

\begin{figure}[h!]
    \centering
    \begin{minipage}[b]{0.7\linewidth}
        \centering
        \includegraphics[width=1.0\linewidth]{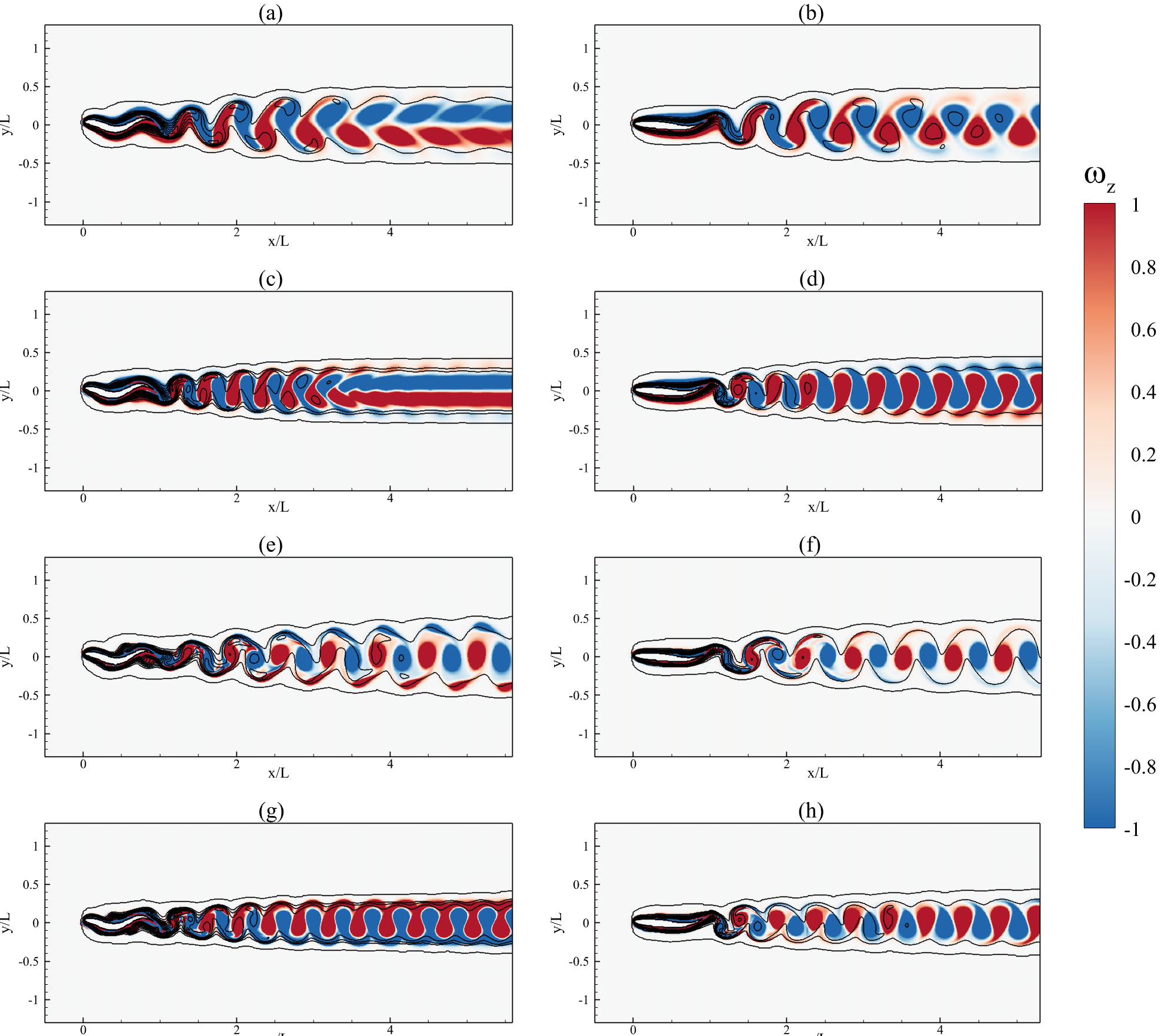}
    \end{minipage}
    \caption{\small \justifying Overlay plots for vorticity and odor concentration are presented for anguilliform and carangiform swimmers in the first and second columns, respectively. Panels (a), (b) and (c), (d) display scenarios with a Re = 1000 and $f^\ast$ = 0.3 and 0.5, respectively. Panels (e), (f) and (g), (h) display scenarios with a Re = 5000 and $f^\ast$ = 0.3 and 0.5, respectively. }
    \label{centrecontour}
\end{figure}

\begin{figure}[h!]
    \centering
    \begin{minipage}[b]{0.65\linewidth}
        \centering
        \includegraphics[width=1.0\linewidth]{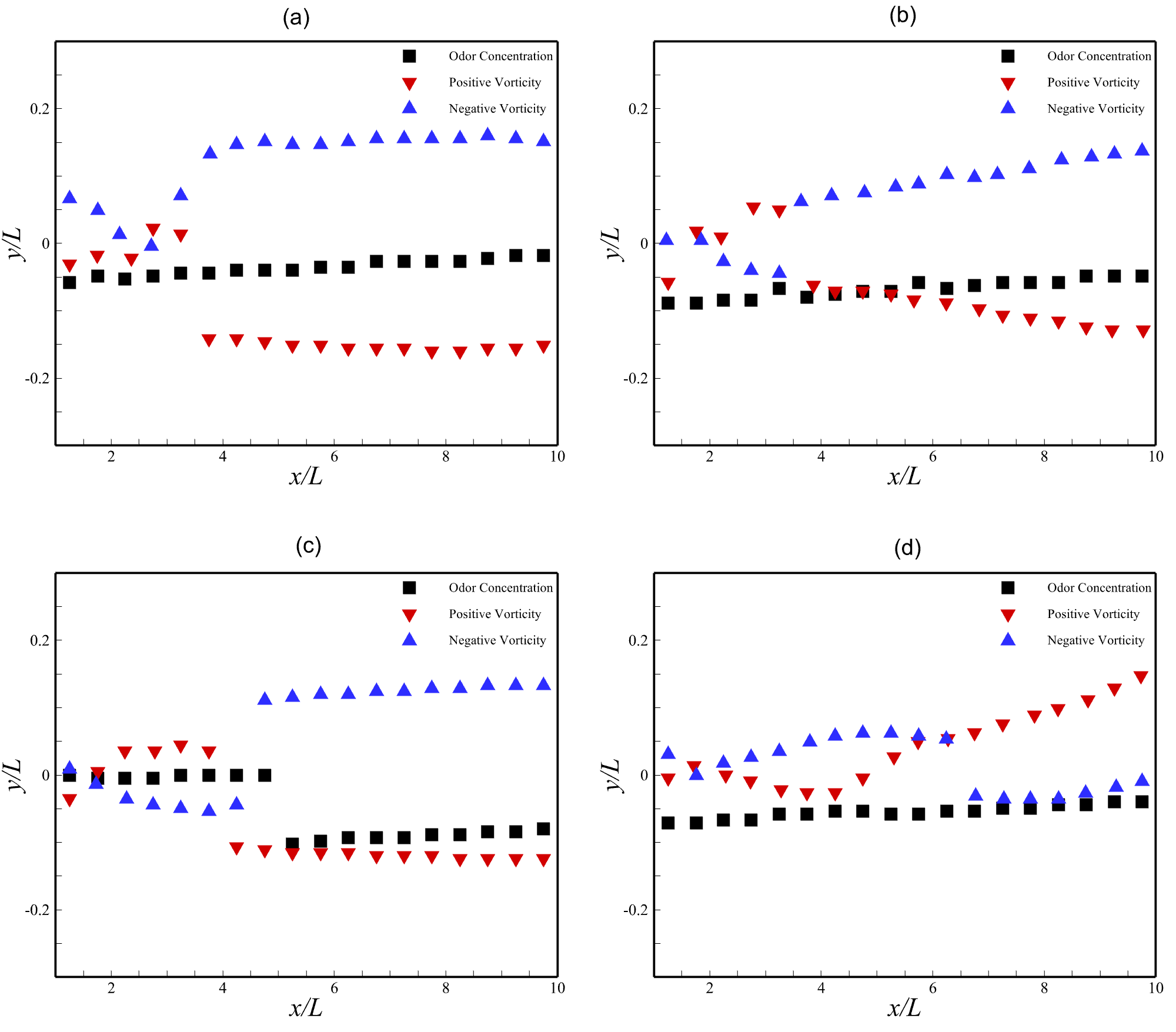}
    \end{minipage}
    \caption{\small\justifying {Plot for coordinates of the center of vortices and odor spots for undulating swimmers are displayed in the first and second columns for anguilliform and carangiform swimmers, respectively. Panels (a) and (b) correspond to Re = 1000 and $f^\ast = 0.3$, while panels (c) and (d) correspond to $f^\ast = 0.5$.}}
    \label{centreplot}
\end{figure}

Another observation in the plots of Figs.~\ref{ov_angulli} and \ref{ov_carangi} corresponds to strong odor spots quantified as the magnitude of $C^\ast$ for the higher $f^\ast$ for both $\mbox{Re}$. We find that odor is transported more strongly and to a farther distance in the wake of the undulating swimmers if they swim with a higher Strouhal frequency, irrespective of the Reynolds number. Based on the magnitude of $C^\ast$, a comparative look at the two sets of plots in Figs.~\ref{ov_angulli} and \ref{ov_carangi} clearly indicate that the anguilliform swimmers is able to spread the odor more strongly to a farther distance in its wake, despite the fact that the carangiform swimmer undulates its body with a higher wave-speed. } 

{After establishing a broad connectivity between odor and vorticity fields around the two undulating swimmers, the next important aspect is that how closely transport of chemical cues is linked with the dynamics of coherent flow structures. To explore this aspect, we plot the contours of the two parameters, $\omega_z$ and $C^\ast$, overlaid on each other in Figs.~\ref{centrecontour}a$-$\ref{centrecontour}h. The plots of $C^\ast$ are shown in the form of solid black lines in all the cases here. It is evident that the not only the overall trajectory of the odor sports are driven by vortices, but also the periphery of the region with high vorticity is overlapped by that of the odor cues. Nevertheless, a careful look at inside the zones of higher activity of vorticity and odor transport in each case, the odor spots very rarely follow the exact paths of the vortices. It means that the odor spots are misaligned with the coherent flow features.}

 {A better way to further analyze this behavior can be to plot the x- and y-coordinates of the centers of vortices and odor spots that is provided in Fig.~\ref{centreplot}. Employing the methodology introduced and explained by Khalid et al. \cite{khalid2020flow}, we compute these coordinates (X, Y) of the centers of vortices and spots of odor through a local search to identify the locations of maximum and minimum vorticity and maximum odor concentration. In all the four cases presented here for different swimming modes and $f^\ast$ at $\mbox{Re}=1,000$, we do not observe a strongly linked local kinematics of chemical cues with either positive or negative vortices. The centers of the odor spots do not follow the cores of the vortices while traversing in the wake even when they are tracked to a distance of $10c$ from the trailing edge of the foil.} 

{The next important element of this work is to determine the influence of convection and diffusion of chemical cues in the spread of odor in the wake.} The interplay between diffusion and convection is {of fundamental significance} in shaping the dispersion and detection of chemical cues in aquatic environments.

\begin{figure}[h!]
    \centering
    \begin{minipage}[b]{1\linewidth}
        \centering
        \includegraphics[width=\linewidth]{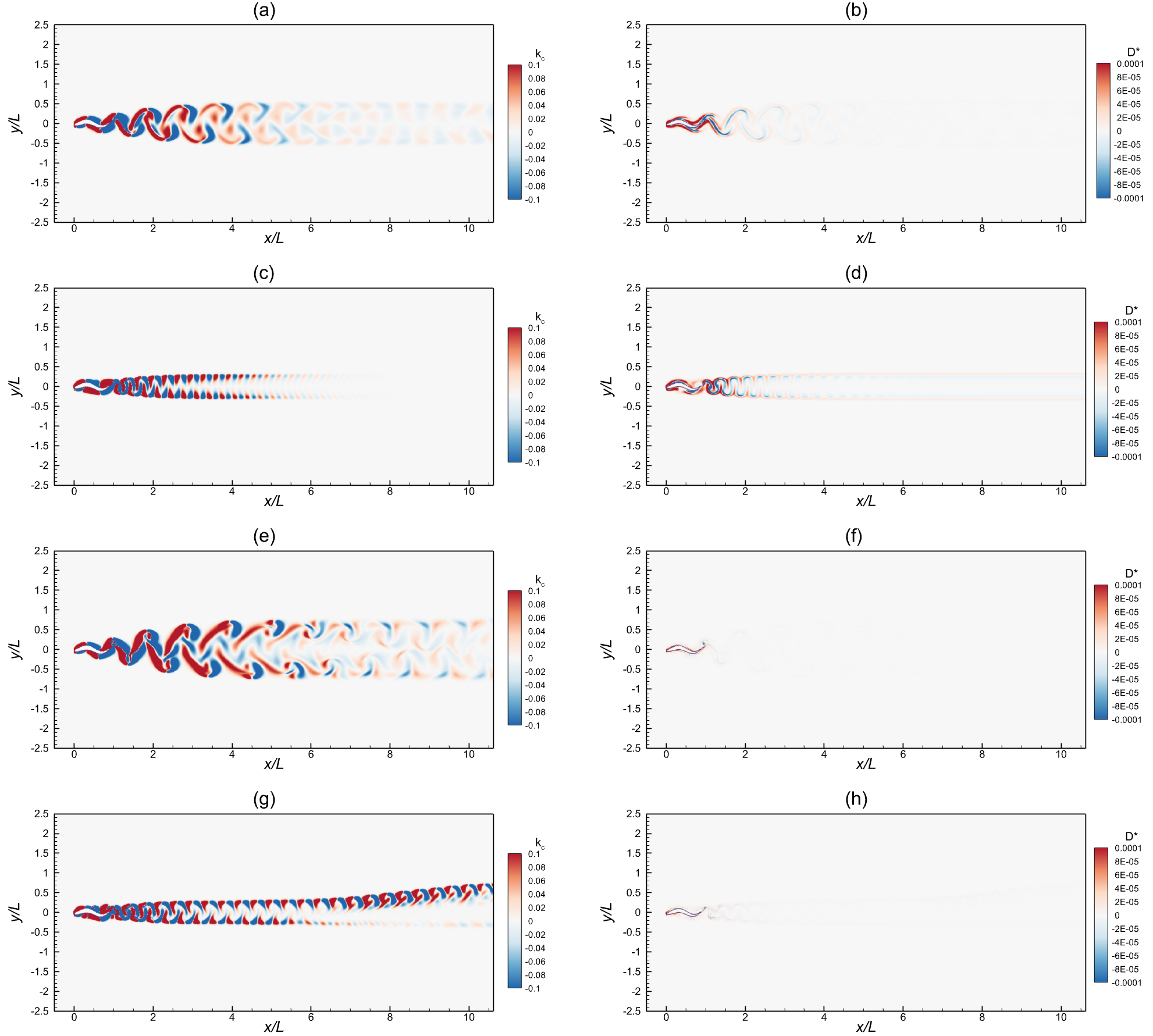}
    \end{minipage}
    \caption{\small\justifying Odor dynamics for anguilliform swimmers, with the first and second columns corresponding to convection and diffusion, respectively. Panels (a), (b) and (c), (d) display scenarios with a Re = 500 and $f^\ast = 0.2$ and $0.6$, respectively. Panels (e), (f) and (g), (h) display scenarios with a Re = 5000 and $f^\ast = 0.2$ and $0.6$, respectively.}
    \label{cd_angulli}
\end{figure}

\begin{figure}[h!]
    \centering
    \begin{minipage}[b]{1\linewidth}
        \centering
        \includegraphics[width=\linewidth]{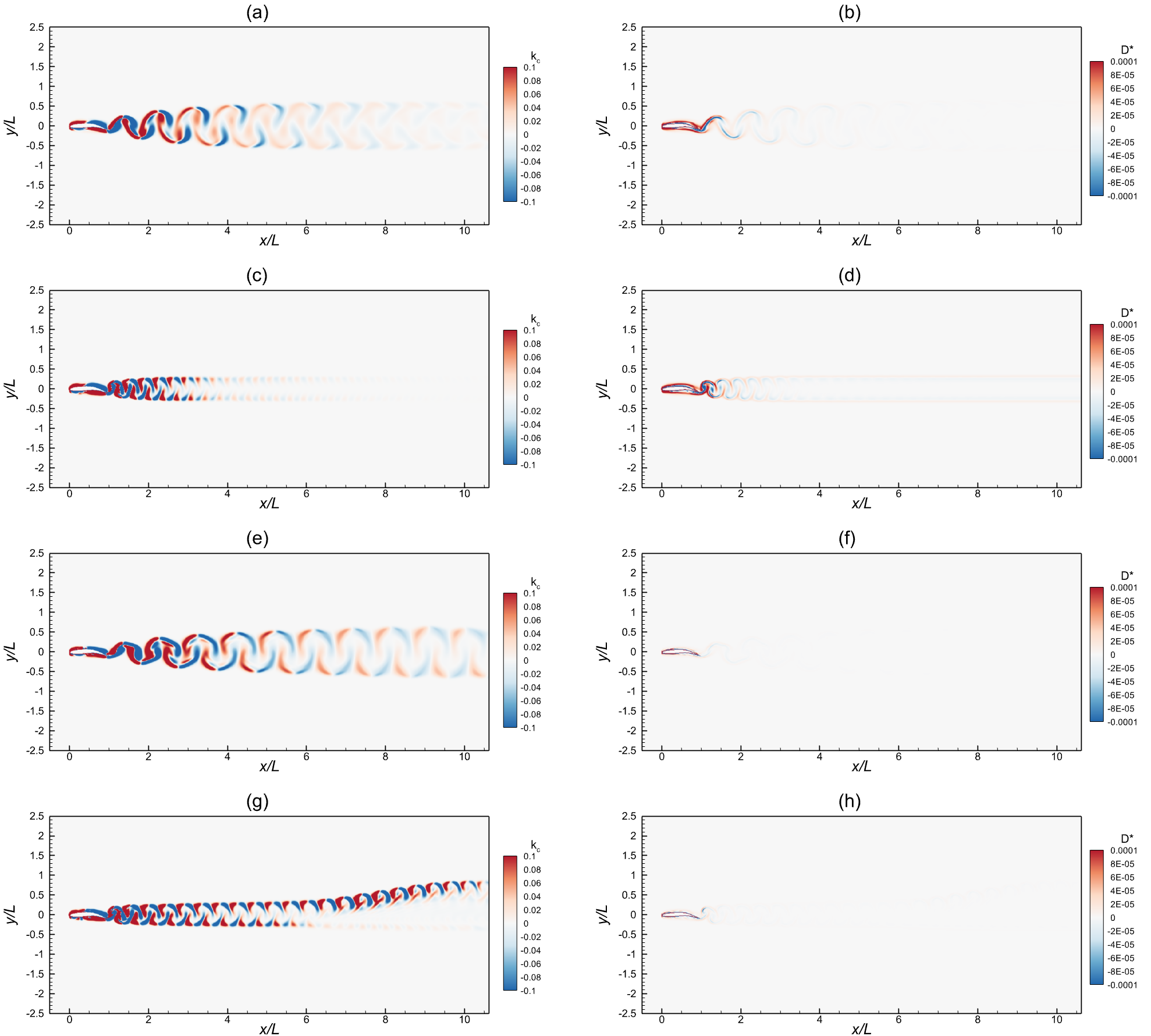}
    \end{minipage}
    \caption{ \small\justifying Odor dynamics for carangiform swimmers, with the first and second columns corresponding to convection and diffusion, respectively. Panels (a), (b) and (c), (d) display scenarios with a Re = 500 and $f^\ast$ = 0.2 and 0.6, respectively. Panels (e), (f) and (g), (h) display scenarios with a Re = 5000 and $f^\ast$ = 0.2 and 0.6, respectively.}
    \label{cd_carangi}
\end{figure}

To evaluate the relative influence of these two distinct phenomena on concentration and transport of odor, we employ the unsteady odorant advection-diffusion equation (Eq.~\ref{eqn:odor_transport}) to evaluate their individual contributions. {For $Sc=340$ for water, our observations from Fig.~\ref{cd_angulli} for the anguilliform swimmer and Fig.~\ref{cd_carangi} for the carangiform swimmer, identify that as the Reynolds number increases from $500$ to $5,000$, convection becomes increasingly dominant in spreading odorants, thereby enhancing the dispersion of chemical signals in the wake. This rise in the convective process improves the mixing and distribution of odor molecules, allowing for more effective and widespread detection of chemical signals far from the body. These observations do not seem to be influenced by the $f^\ast$ and the type of wave-form imposed on the swimmer's body. The plots in the right-sided columns of Figs.~\ref{cd_angulli} and \ref{cd_carangi} demonstrate that the diffusion process significantly diminishes with the increase of $\mbox{Re}$, but a higher $f^\ast$ slightly improves its strength in the wake.} 

\begin{figure}[h!]
    \centering
    \begin{minipage}[b]{1\linewidth}
        \centering
        \includegraphics[width=\linewidth]{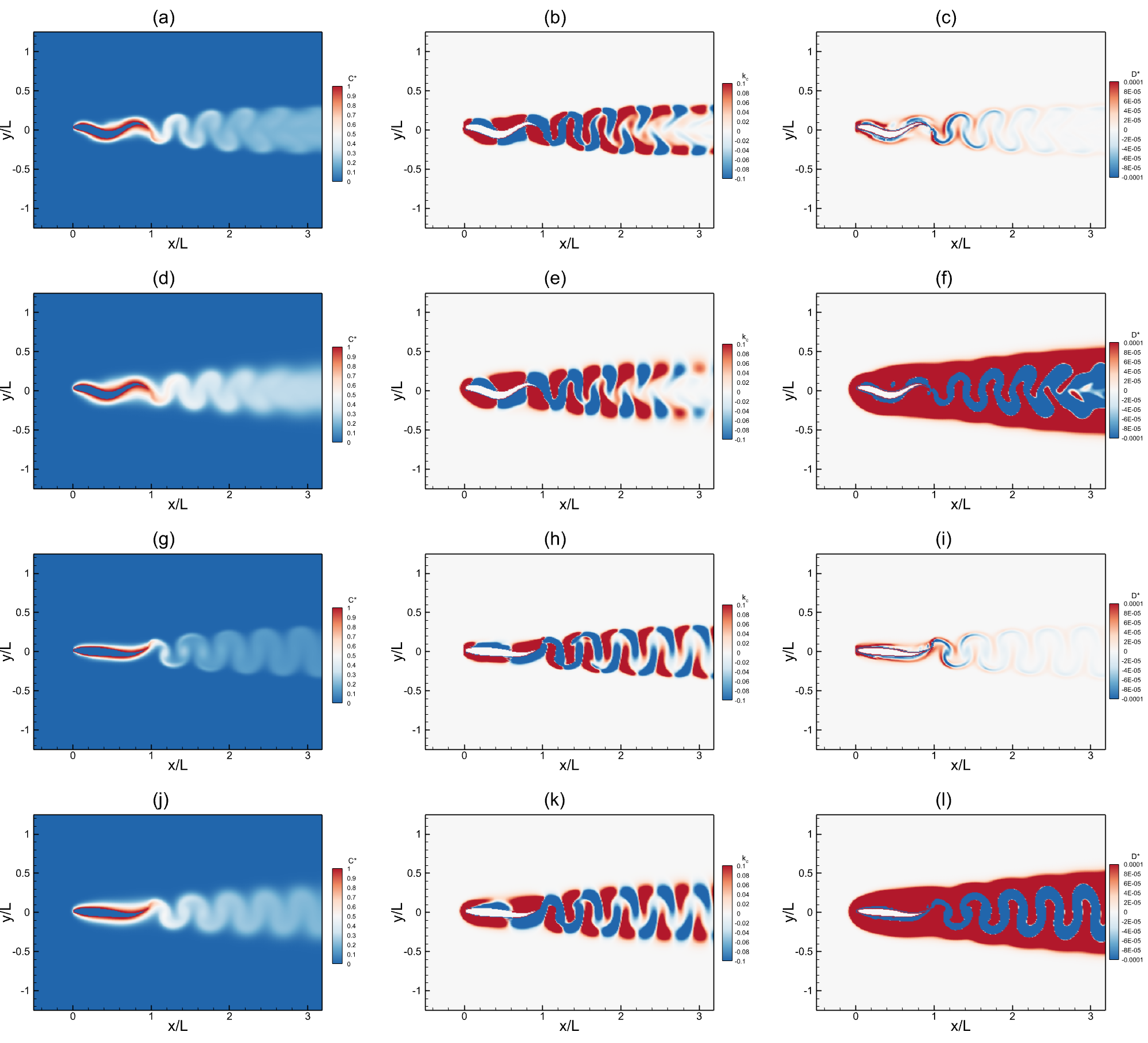}
    \end{minipage}
    \caption{ \justifying Side-by-side comparison of odorant dynamics, focusing on concentration (first column), convection (second column), and diffusion (third column). Panels (a) through (f) and (g) through (l) represent anguilliform and carangiform swimmers, respectively. Panels (a) to (c) and (g) to (i) show water as the medium, while panels (d) to (f) and (j) to (l) depict air as the medium.}
    \label{schmidt_a_c}
\end{figure}

{Because most of the previous studies in literature \cite{li2018balance, lei2023numerical, lei2023wings} focussed on the phenomenon relevant to $Sc$ ranging upto $100$, which does not correspond to the conditions in water, we make it an integral part of our present work due to the importance of fluid-structure-chemical interactions in marine environments. Although undulatory motion belongs more to fish species propelling in water, we test and demonstrate} the robustness of our in-house solver by {considering $Sc$ for two media $-$ water and air $—$ to elaborate how convection and diffusion processes are impacted by the Schmidt number and how this parameter determines the dominance of one phenomenon over the other. For this purpose, we choose} a transitional flow regime with $f^\ast$ of $0.45$ for both undulating swimmers {as representative cases in Fig.~\ref{schmidt_a_c}}. In aquatic environments like water, where the value $Sc$ is high for water, convection predominantly drives the dispersion of chemical cues. The high density and viscosity of water enhance turbulent mixing, thereby promoting effective convective transport while diminishing the relative impact of molecular diffusion. Conversely, $Sc$ decreases substantially in air due to its lower density and viscosity, making molecular diffusion more significant in odorant distributions, as illustrated in Figure \ref{schmidt_a_c}. The reduced convective mixing in the air allows diffusion to play a more prominent role in the spread of chemical signals, leading to a more gradual and extensive dispersion pattern. We observed that in air, the impact of odor concentration is stronger due to the additional effect of diffusion. {On the other hand}, diffusion has a minimal effect on odor concentration {in water}, with convection being the primary mechanism for odor dispersion. Therefore, in aquatic environments, the spread of odor is primarily driven by convection, while in air, both convection and diffusion {play significant roles, and contributions from none of them may be disregarded.}

\section{Conclusions}
In this paper, we conduct a comprehensive analysis on the interactions between odor dynamics and vortex flow around undulating bodies in fluid environments, offering valuable insights into bio-inspired propulsion and sensing mechanisms based on transport of chemical cues. We investigat how specifically carangiform and anguilliform kinematic modes influence the dispersion of odorants in both water and air environments. Our findings show that while odor fields are generally linked with vortex dynamics, odor spots do not consistently follow the exact paths of vortices. Although odor spots tend to overlap with regions of high vorticity, they are often misaligned with the coherent flow structures. The centers of odor spots do not consistently align with vortex cores, even when tracked over large distances from the source, indicating a complex and non-linear relationship between odor transport and vortex dynamics. At higher $f^\ast$, odor is transported more strongly and over greater distances in the wake, with anguilliform swimmers spreading odor more effectively than carangiform swimmers. Flow conditions and kinematic parameters, such as Reynolds number and Strouhal frequency, further influence the interconnected dynamics of fluid flow and chemical cue transport. Higher Reynolds numbers and Strouhal numbers enhance convective transport, resulting in more effective odor dispersion. We find that, in water with its higher Schmidt number, convection is the dominant mechanism for odor dispersion, while in air, diffusion plays a more significant role, leading to a slower and broader spread of chemical cues. Overall, our study advances the understanding of the interplay between fluid dynamics and chemosensory processes in aquatic locomotion, offering important insights for the development of efficient bio-inspired robotic systems capable of operating effectively in diverse environmental conditions. The integration of odor dynamics with vortex dynamics in fluid flow represents a novel contribution to the field, with potential applications in designing advanced underwater robotic systems capable of navigating complex marine environments.

\section*{Acknowledgement}
MSU Khalid acknowledges funding support from the Natural Sciences and Engineering Research Council of Canada (NSERC) through the Discovery and Alliance International grant programs for this work. C. Li acknowledges funding support from the National Science Foundation (CBET-2042368) monitored by Dr. R. D. Joslin and Air Force Office of Scientific Research (FA9550-24-1-0122) monitored by Dr. Patrick Bradshaw.

\vskip6pt









\bibliography{references}



\end{document}